\documentclass[a4paper,11pt]{article}
\pdfoutput=1 % if your are submitting a pdflatex (i.e. if you have images in pdf, png or jpg format)

\usepackage{jheppub}

\usepackage[ddmmyyyy,hhmmss]{datetime}

\usepackage{amsmath}
\usepackage{amsfonts}
\usepackage[retainorgcmds]{IEEEtrantools}
\usepackage{tikz-feynman}
\usepackage{float}
\usepackage{verbatim}
\usepackage{slashed}
\usepackage{empheq}
\usepackage[normalem]{ulem}

\hypersetup{pdfstartview={XYZ null null 0.75}}
\allowdisplaybreaks

\makeatletter \@addtoreset{equation}{section} \makeatother

\newcommand{\OO}{\mathcal{O}}
\newcommand{\LL}{\mathcal{L}}

\newcommand{\N}{\mathcal{N}}

\newcommand{\G}{\mathcal{G}}

\newcommand{\K}{\mathcal{K}}

\usepackage{amsmath,graphicx,amssymb,subfigure,bbm,comment}
\usepackage[all]{xy}
\usepackage{mathtools}
\usepackage{amsfonts}
\usepackage{comment}
\usepackage{mdframed}

\usepackage{feynmp}
\tikzfeynmanset{compat=1.1.0}

\usepackage{xcolor}
\usepackage{pifont}

\allowdisplaybreaks

\def\be{\begin{equation}}
\def\ee{\end{equation}}
\def\ba{\begin{eqnarray}}
\def\ea{\end{eqnarray}}

\def\a{\alpha}

\def\b{\beta}

\def\g{\gamma}
\def\G{\Gamma}
\def\d{\delta}
\def\D{\Delta}

\def\l{\lambda}
\def\L{\Lambda}
\def\s{\sigma} 
\def\S{\Sigma}

\def\cN{{\cal N}}

\def\IR{\relax{\rm I\kern-.18em R}}

\def\IR{\relax{\rm I\kern-.18em R}}
\def\inv{^{\raise.15ex\hbox{${\scriptscriptstyle -}$}\kern-.05em 1}}

\def\vev#1{{\left<#1\right>}}

\title{%Interpolating holographic defect CFTs\\
\huge Weyl Anomaly Coefficients of Holographic Defect CFTs at Weak and Strong Coupling}

\author{\Large George Georgiou}

\emailAdd{georgios.georgiou2@gmail.com}
\abstract{We determine the type-A Weyl anomaly coefficient $b$, associated with the  intrinsic scalar curvature of the defect,  for the class of holographically realised co-dimension two defect CFTs (dCFTs) introduced in \cite{Georgiou:2025mgg}  and  \cite{Georgiou:2025wbg}.  At strong coupling, we employ the dual D5-brane solutions in Euclidean signature, where the defect is supported on an $S^2$ submanifold of the Euclidean $AdS_3\times S^1$ boundary. At weak coupling, we use the classical solutions of the ${\cal N}=4$ SYM equations of motion, previously conjectured to describe the defects dual to the D5-brane configurations. Notably, the coefficient $b$ is found to be negative in a finite region of parameter space. To our knowledge, this constitutes the first explicit example of an {\it interacting} unitary dCFT with $b<0$. 

We also compute the type-B Weyl anomaly coefficients associated with the extrinsic curvature of the defects, first at strong coupling and subsequently at weak coupling. In a certain limit, we find agreement between the weak- and strong-coupling results for both the type-A and type-B anomaly coefficients.
}

\begin{document}
\maketitle
\flushbottom

%%%%%%%%%%%%%%%%%%%%%%%%%%%%%%%%%%%%%%%%%%%%%%%%%%%%%%%

\section{Introduction}
%The dynamics of conformal field theories (CFTs) in the presence of defects describes a plethora of physical systems, ranging from boundaries and interfaces to objects such as Wilson loops, strings and branes. In general, the presence of a defect breaks, partially or even completely, the conformal symmetry of the ambient CFT rendering the calculation of observables much more involved. The situation becomes even more intricate if one is interested in the behaviour of the theory in the strong coupling regime. The calculations become rather tractable in the case where the ambient CFT is the maximally supersymmetric gauge theory in 4 dimensions, i.e ${\cal N}=4$ SYM. The reason is two-fold. On one hand, the ambient CFT is believed to be integrable in the planar limit \cite{Minahan:2002ve,Bena:2003wd} and on the other because of its holographic description in terms of type IIB string theory on $AdS_5\times S^5$ \cite{Maldacena:1997re}.

Defect conformal field theories (dCFTs) provide a rich and versatile framework for exploring the interplay between local quantum degrees of freedom and lower-dimensional impurities, boundaries, or interfaces embedded within a higher-dimensional ambient theory. Such systems arise naturally across a wide range of contexts, from critical phenomena and condensed matter systems with impurities to high-energy constructions involving branes and dualities. A central feature of conformal field theories is the presence of trace anomalies, which encode universal information about the theory and serve as powerful probes of its structure. In the presence of defects, these anomalies acquire new contributions localised on the defect, leading to a refined set of observables known as defect Weyl anomaly coefficients.

Understanding these coefficients is important both conceptually and practically. They characterise intrinsic properties of the defect degrees of freedom, govern responses to geometric deformations, and often obey nontrivial constraints such as monotonicity along renormalisation group flows \cite{Jensen:2015swa}. Early investigations of boundary and defect anomalies highlighted the emergence of new central charges associated with lower-dimensional submanifolds \cite{Cardy:1984bb,McAvity:1993ue}, while more recent work has clarified their classification and relation to correlation functions and displacement operators  \cite{Herzog:2017xha,Jensen:2015swa,Billo:2013jda}. Despite this progress, explicit computations remain challenging, especially in strongly coupled regimes.

Holography offers a powerful complementary approach to this problem. Within the gauge/gravity duality, defect CFTs are typically realised by introducing probe or backreacting branes in asymptotically Anti-de Sitter (AdS) spacetimes, thereby providing a geometric encoding of defect degrees of freedom. Foundational work on holographic renormalization and anomaly extraction  \cite{Henningson:1998gx,deHaro:2000vlm,Balasubramanian:1999re} established the tools necessary to compute Weyl anomalies from bulk gravitational actions, while subsequent studies extended these ideas to systems with boundaries and defects \cite{McAvity:1995zd,Takayanagi:2011zk,Fujita:2011fp,Herzog:2017xha,Herzog:2017xha_disp}. In this framework, Weyl anomaly coefficients can be extracted from the gravitational action through holographic renormalisation, translating field-theoretic questions into problems in classical gravity.

Surface operators are extended defects supported on submanifolds of spacetime and play a central role in modern gauge theory, particularly in the study of dualities and non-perturbative phenomena. In four-dimensional supersymmetric field theories, co-dimension two surface operators, supported on two-dimensional loci, provide a natural higher dimensional generalisation of Wilson and 't Hooft line operators. They are defined by imposing singular boundary conditions on the fields along the defect, thereby encoding localised, intrinsically two-dimensional degrees of freedom.
In ${\cal N}=4$ supersymmetric Yang-Mills (SYM) theory, such operators may preserve part of the underlying supersymmetry or break it entirely \cite{Georgiou:2025mgg,Georgiou:2025wbg}. \footnote{For a recent review focusing on non-supersymmetric surface operators see \cite{Georgiou:2026zhf}.}Their significance is amplified by the highly constrained structure of the theory and its exact S-duality symmetry. Typically, surface operators are characterised by prescribed singular profiles for the gauge and scalar fields in the vicinity of the defect \cite{Gukov:2006jk,Gaiotto:2008sa}. They can be constructed either by coupling the four-dimensional theory to a two-dimensional sigma model localized on the defect, or by descending from the six-dimensional $(2,0)$ theory via dimensional reduction \cite{Kapustin:2006pk,Witten:2009at}.
Beyond their intrinsic field-theoretic interest, surface operators in ${\cal N}=4$ SYM are deeply connected to geometric representation theory and the geometric Langlands program. In particular, in the work of Gukov and Witten they arise as data specifying ramification in the geometric Langlands correspondence \cite{Gukov:2006jk,Gaiotto:2009hg}. Under S-duality, these operators transform in a highly non-trivial manner, intertwining electric and magnetic descriptions and relating weak and strong coupling regimes \cite{Gukov:2006jk}.

Within the framework of the AdS/CFT correspondence, surface operators admit a natural realization in terms of extended objects in the bulk string theory. They are typically described by probe branes wrapping appropriate submanifolds of the $AdS_5 \times S^5$ background. For co-dimension two defects, the corresponding holographic duals involve brane configurations whose worldvolumes contain an $AdS_3$ factor-reflecting the conformal symmetry preserved by the defect-and wrap internal cycles in the compact space. Depending on the regime, these branes may either be treated in the probe approximation or fully backreact to produce supergravity solutions with localised $AdS_3$ geometries.
Several canonical examples illustrate these constructions. Although of co-dimension one, the D3-D5 brane intersection \cite{Karch:2000gx} provides a useful prototype, realising a defect theory with ${\cal N}=(4,4)$ supersymmetry. A more direct example is the $\tfrac{1}{2}$-BPS D3-D3 intersection, whose gravity dual is described by a probe D3-brane wrapping $AdS_3 \times S^1$ inside $AdS_5 \times S^5$ \cite{Drukker:2008wr}, with the fully backreacted geometry constructed in \cite{Gomis:2007fi}. More general surface operators -featuring reduced supersymmetry or additional structures such as monodromies, theta angles, or two-dimensional gauge fields- can be engineered via intersecting brane configurations or through M2-branes ending on M5-branes in M-theory. These realisations establish a precise correspondence between field-theoretic data, such as the singular behaviour of fields, and geometric features in the dual string description, including brane embeddings and worldvolume fluxes.

In this paper, we investigate the Weyl anomaly coefficients associated with holographically realised defect CFTs. We systematically compute defect contributions to the anomaly from the bulk perspective, carefully accounting for  the embedding geometry of the defect. Our analysis applies to a broad class of holographic models which are realised through probe brane setups. We further explore how these coefficients depend on geometric and physical data, and we compare our results with the field-theoretic expectations and constraints.

The structure of the paper is as follows. In section \ref{anomal}, we review the general structure of Weyl anomalies in the presence of defects and introduce the relevant geometric invariants, as well as the way to determine the type-A and type-B anomaly coefficients of the defects. 
Section \ref{anom-sol-0} outlines the calculation of the type-A defect anomaly coefficient for the co-dimension 2 defect CFT of \cite{Georgiou:2025mgg}. The calculation is performed in Euclidean signature with the locus of the defect being a two-sphere $S^2$. First we perform the holographic calculation of the anomaly coefficient $b$ by the use of the dual to the dCFT D5 brane. Subsequently, we evaluate the same coefficient at weak coupling by employing the classical solution of the ${\cal N}=4$ SYM equations of motion which describe the  defect in the field theory picture \cite{Georgiou:2025mgg}. At a certain limit perfect agreement between the weak and strong coupling results for $b$ is found.
In section \ref{anom-sol-1}, we present the analogous  computations of the defect anomaly coefficient $b$ for the generalised duality of \cite{Georgiou:2025wbg}. Section \ref{anom-sol-d1} is devoted to the calculation of the type-B Weyl anomaly coefficient $d_1$ that is associated with the extrinsic curvature term for the holographically realised co-dimension 2 defect CFT of \cite{Georgiou:2025wbg}. At first we perform the holographic calculation and then 
the calculation at the weak coupling regime. As for the type-A Weyl anomaly coefficient $b$, agreement is observed in the appropriate limit. The anomaly coefficient for the dCFT of \cite{Georgiou:2025mgg} is obtained as the appropriate  limit of the result for the dCFT of \cite{Georgiou:2025wbg}.
Finally, in section \ref{concl}, we draw our conclusions.%discuss the implications of our results and possible directions for future work.

\section{Defect anomalies}\label{anomal}

Here we consider a two-dimensional defect $\S$ embedded in a $d$-dimensional, generically curved, manifold with metric $g_{\mu\nu}$. The defect is parametrised by $\s^a,\,a=1,2$ with its embedding  in the $d$-dimensional space described by $X^\mu(\s),\, m=0,1,\cdots d-1$.

Let us now  focus on the renormalised partition function $Z=e^{-W}$ as a functional of the metric $g_{\mu\nu}$, the embedding coordinates $X^{\mu}$ and the set of all marginal and relevant couplings, $\{\lambda\}$. The one-point functions of the stress-energy tensor $\langle T^{\mu\nu}\rangle$ and the displacement $ \langle D_{\mu} \rangle$  can be derived from the variation of $W \equiv -\ln Z[g_{\mu\nu},X^{\mu},\{\lambda\}]$ with respect to $g_{\mu\nu}$ and $X^{\mu}$, respectively
\begin{align}
\label{E:deltaW}
\delta W = -\frac{1}{2}\int d^dx \sqrt{g}\, \delta g_{\mu\nu}\langle T_{\textrm{b}}^{\mu\nu} \rangle
%\\
%\nonumber
\quad -\int d^2\sigma \sqrt{\g} \left[ \frac{1}{2}\delta g_{\mu\nu} \langle T_{\textrm{d}}^{\mu\nu}\rangle +  \delta X^{\mu} \langle D_{\mu}\rangle  + \hdots\right]\, .
\end{align}
Here $g$ and $\g$ denote the determinants of the bulk $g_{\mu\nu}$ and the induced on the defect $\g_{ab}$ metrics, respectively. The dots indicate terms involving derivatives of $\delta g_{\mu\nu}$ normal to the defect which may or may not be present. By rewriting the defect's volume as $\int d^2\sigma \sqrt{\g}=\int d^dx \sqrt{g} \, \delta^{d-2}(x_\bot)$ one can see  that $\langle T^{\mu\nu} \rangle$ receives contributions both from the  bulk and the defect.  The subscripts b and d indicate those contributions
\be\label{S-ET}
\langle T^{\mu\nu}\rangle = \langle T^{\mu\nu}_{\textrm{b}}\rangle + \delta^{d-2}(x_\bot) \langle T^{\mu\nu}_{\textrm{d}}\rangle + \hdots\, .
\ee
The dots in \eqref{S-ET} indicate terms involving normal derivatives of $\delta^{d-2}(x_\bot)$ which are coming from the corresponding dots in~\eqref{E:deltaW}. Higher-order variations of $W$ give, as usual, the connected higher-point correlation functions.

One may now perform an infinitesimal Weyl variation defined by
\be\label{Weyl-var}
\delta_{\omega} g_{\mu\nu}=2\omega g_{\mu\nu}, \qquad \delta_{\omega} X^{\mu}=0,
\ee
where $\omega$ is the parameter of the variation.
Conformal field theories are Weyl-invariant except for a potential  anomalous term. Namely, the variation of $W$  under an infinitesimal Weyl variation will be generically given by
\be
\label{E:WeylAnomaly}
\delta_{\omega} W = -\int d^dx \sqrt{g}\, \omega \, \mathcal{A}\, .
\ee
The local function $\mathcal{A}$ is built from the metric $g_{\mu\nu}$, the embedding coordinates $X^\mu$ and the pullback of the Weyl tensor on the defect. %Indeed, we will only consider contributions to $\mathcal{A}$ built from $g_{\mu\nu}$ alone.
Plugging \eqref{Weyl-var} in \eqref{E:deltaW} and comparing to ~\eqref{E:WeylAnomaly} leads to the Weyl Ward identity
\be
\langle T^{\mu}_{~\mu}\rangle = \mathcal{A}\,. 
\ee
The  form of the anomaly $\mathcal{A}$ can be determined by solving the Wess-Zumino (WZ) consistency condition~\cite{Wess:1971yu}.
Since  the Weyl group is Abelian this reduces to the requirement that 
two successive Weyl transformations of the functional $W$ should commute. For a CFT living in even dimensions \footnote{For odd $d$ the bulk contribution to $\mathcal{A}=0$~\cite{Deser_1993}.}, $d=2 n$,  the solution of WZ consistency condition reads~\cite{Deser_1993}
\be
\label{E:bulkAnomaly}
\mathcal{A}_{\textrm{b}} = (-1)^{\frac{d}{2}+1}\frac{4a}{d!\text{vol}(\mathbb{S}^d)}E_d + \sum_I c_I W_I,
\ee
where  $E_d$ is the Euler density and  $W_I$ are the Weyl-covariant scalars of weight $-d$. At this point let us mention that WZ consistency allows also for total derivatives in~\eqref{E:bulkAnomaly}, which can be eliminated by the use of  local counterterms. The coefficients $a$ and the $c_I$ are quantities that are left undetermined by the WZ consistency conditions. 

In the case of a defect CFT, as it happens with the stress-energy tensor, the anomaly $\mathcal{A}$ receives distinct contributions from both the bulk and the defect namely, $\mathcal{A} = \mathcal{A}_{\textrm{b}} + \delta^{d-2}(x_\bot)  \mathcal{A}_{\textrm{d}}$. The bulk term $\mathcal{A}_{\textrm{b}}$ is given by \eqref{E:bulkAnomaly}. 
In the case of a two-dimensional defect in $d\geq 3$ the solution of the WZ consistency condition gives  for the defect anomaly term, $\mathcal{A}_{\textrm{d}}$, the following expression \cite{Henningson:1999xi,Schwimmer:2008yh},
%\be
%\label{E:defectAnomaly}
%\mathcal{A}_{\textrm{d}} = \frac{1}{24\pi}\left( b \, \hat{R}+ d_1 \, \mathring{\II}^{\mu}_{ab}\mathring{\II}_{\mu}^{ab} + d_2 \, W_{abcd} \hat{g}^{ac}\hat{g}^{bd}\right),
%\ee
%with $\hat{R}$ the Ricci scalar of $\hat{g}_{ab}$, $\mathring{\II}^{\mu}_{ab}$ 
%the traceless part of $\II^{\mu}_{ab}$, 
%and $W_{abcd}$ the pullback of the bulk Weyl tensor. WZ consistency leaves undetermined the ``defect central charges'' $b$, $d_1$, and $d_2$. (The Weyl tensor vanishes identically in $d=3$, so $d_2$ exists only for $d \geq 4$.)
\be
\label{eq:defect-A}
\langle T^{\mu}_{~\mu} \rangle_{\textrm{d}} =\frac{1}{24\pi}\left( b \, {\cal{R}}_{def} + d_1 \, Y^{\mu}_{ab}Y_{\mu}^{ab} - d_2 \, W_{ab}^{~~ab} \right),
\ee
where ${\cal{R}}_{def}$ denotes the intrinsic scalar curvature of the defect  while $W_{abcd}$ is the pullback of the bulk Weyl tensor to the defect. Moreover, 
\be\label{Y}
Y^{\mu}_{ab}\equiv \Pi^{\mu}_{~ab} - \frac{1}{2}\gamma_{ab} \gamma^{cd}\Pi^{\mu}_{~cd}
\ee
is the traceless part of the second fundamental form which is defined through $\Pi^{\mu}_{~ab} = \hat{\nabla}_a \partial_b X^{\mu}$.
The bulk covariant derivative $\nabla_{\mu}$ induces a defect covariant derivative $\hat{\nabla}_a$. The latter can act on tensors which carry  both bulk and along the defect indices as follows: 
\be
\hat{\nabla}_a M^{\mu}_b \equiv \partial_a M^{\mu}_b + \Gamma^{\mu}{}_{\nu a} M^{\nu}_b - \hat{\Gamma}^c{}_{ab} M^{\mu}_c\,,
\ee
where $\Gamma^{\mu}_{~\nu a}$ is the pullback of the Levi-Civita connection
\be
\Gamma^{\mu}_{~\nu a} = \Gamma^{\mu}{}_{\nu\rho} \partial_a X^{\rho}\, .
\ee
Moreover,  $\hat{\Gamma}^a_{~bc}$ is the Levi-Civita connection derived from the induced on the defect metric $\g_{ab}$. The action of $\hat{\nabla}_a$ on more general tensors should be obvious.
As can be seen from \eqref{eq:defect-A}, the defect is characterised by the anomaly coefficient $b$, $d_1$, and $d_2$ which are the defect central charges.

Under a Weyl transformation, $\sqrt{\g} \, {\cal{R}}_{def}$ transforms as a total derivative (in the classification of~\cite{Deser_1993} this term is of type A ), while 
$\sqrt{\g}\,Y^{\mu}_{ab}Y_{\mu}^{ab} $
and $\sqrt{\g} \, W_{ab}^{~~ab}$ are  Weyl-invariant (that is of type B). The  anomaly coefficient $b$ is the analogue of the bulk central charge $a$, which obeys the $c$ or $a$ theorem in $d=2$ or $4$, respectively, while the coefficients $d_1$ and $d_2$ correspond to the $c_I$ of \eqref{E:bulkAnomaly}.

We now describe the strategy in order to calculate the anomaly coefficients $b$, $d_1$ and $d_2$.\\
$\bullet$ 	Put the theory on a curved background with an embedded defect $\S$.\\
$\bullet$ 	Introduce a Weyl rescaling 
\be\label{Weyl-fin}
g_{\mu\nu}\rightarrow e^{2\omega} g_{\mu\nu}.
\ee
\\
$\bullet$ 	Isolate the logarithmic divergence of the regularised effective action: 
\be\label{Wdiv}
W_{div}=-\, \log\frac{a_\S}{\epsilon} \,\int d^2\sigma \sqrt{\g} \left( \tilde b \, {\cal{R}}_{def} + \tilde d_1 \, Y^{\mu}_{ab}Y_{\mu}^{ab} - \tilde d_2 \, W_{ab}^{~~ab} \right)+\cdots
\ee
\\
$\bullet$   Extract $b$, $d_1$ and $d_2$ as the coefficients of the Ricci scalar ${\cal{R}}_{def}$, the extrinsic curvature term $Y^{\mu}_{ab}Y_{\mu}^{ab} $ and the trace of the pullback of the Weyl tensor $ W_{ab}^{~~ab}$, respectively. In \eqref{Wdiv}, $\epsilon$ is a short distance cut-off (which becomes the UV cut-off $\L=\frac{1}{\epsilon}$ in field theory) needed to regularise the effective action, while $a_\S$ is some characteristic scale of the defect which makes the argument of the logarithm dimensionless. Moreover, the dots indicate terms which are proportional $\frac{1}{\epsilon^n}$ and can be removed by adding to the action the appropriate counterterms.

To save space we concentrate  on the term associated with the  extrinsic curvature of the defect and suppress all other terms.
Under the Weyl rescaling \eqref{Weyl-fin}  the regulator $a_\S \rightarrow a_\S \, e^\omega$  while $\sqrt{\g}\,Y^{\mu}_{ab}Y_{\mu}^{ab}$  and the cut-off $\epsilon$ remain invariant, as mentioned above. Consequently the variation of $W_{div}$ becomes
\be\label{dW}
\delta_\omega W_{div}=-\frac{1}{24\pi}\, \omega \,\int d^2\sigma \sqrt{\g}  \,d_1 \, Y^{\mu}_{ab}Y_{\mu}^{ab}  +\cdots
\ee
Comparing equation \eqref{E:WeylAnomaly} with $\omega={\rm const.}$ with the last equation \eqref{dW} we see that the anomaly coefficient $d_1$ can be obtained from \eqref{Wdiv} as the coefficient of the logarithm of the cut-off $\log \epsilon=-\log \L$. More precisely, 
\be\label{coeffs}
b= 24 \pi\, \tilde b,  \quad d_1= 24 \pi\, \tilde d_1, \quad {\rm and} \quad  d_2= 24 \pi\, \tilde d_2\, .
\ee

\section{Type-A defect anomaly coefficient for the defect CFT of \cite{Georgiou:2025mgg}}\label{anom-sol-0}

\subsection{The D5 brane of \cite{Georgiou:2025mgg} in Euclidean signature}\label{D5-sol-1}

In this section  we will rewrite the D5-brane solution of \cite{Georgiou:2025mgg} which realises the gravity duals of certain non-supersymmetric 
co-dimension 2 defect conformal field theories in Euclidean signature. This is convenient because we would like the co-dimension 2 defect to have support on a two-sphere $S^2\subset AdS_3\times S^1 $ of the $AdS_3\times S^1 $ boundary. More precisely, we choose the following coordinate system parametrising the $AdS_5\times S^5 $ type IIB supergravity background
\begin{equation}\label{metric-Eucl}
ds^2 = \cosh^2u\, \Big[d\hat\rho^2 + \sinh^2\hat\rho (d\theta_1^2+\sin^2\theta_1 d\phi^2)
 \Big] +du^2+ \sinh^2u \, d\psi^2 +d\Omega_5^2
\end{equation}
with the length element of $S^5$ being
\begin{equation}\label{metric}
d\Omega_5^2 = 
d\tilde\psi^2 + 
\sin^2 \tilde\psi \, \left(d{\tilde \beta}^2 + \sin^2 {\tilde \beta} \, d{\tilde \gamma}^2 \right) +
\cos^2 \tilde\psi \left(d\beta^2 + \sin^2 \beta \, d\gamma^2\right) \, . 
\end{equation}
In addition, we choose the RR 4-form potential to be given by
\begin{equation}\label{C4}
C_4 = (\cosh^4u-1)  \sinh^2\hat\rho \sin\theta_1 \, d\hat\rho \wedge d\theta_1 \wedge d\phi \wedge d\psi \, . 
\end{equation}
Let us mention that the choice of the 4-form potential is different from \cite{Drukker:2008wr} but the same as in \cite{Jiang:2024wzs} and \cite{Gutperle:2020}. This choice ensures that the value of the $b$ anomaly coefficient as obtained from the 1/2-BPS D3 probe brane agrees with the general expression for the $b$ coefficient found in \cite{Jensen:2018rxu,Chalabi:2020iie}.
The boundary of the space $\partial AdS_5$ is at $u\rightarrow \infty$. Thus, the dual field theory will be the Euclidean version of $\mathcal{N}=4$ SYM living in a spacetime with geometry $AdS_3\times S^1$.

The dynamics of the D5 brane  will be governed by the Euclidean action
\begin{equation}
\label{D5-Eucl}
S_{E}= \frac{T_5}{g_s}\Bigg\{\int d^6 \zeta \sqrt{{\rm det}\, \mathcal P [g+2 \pi \a' F]}-
2 \pi \a' \int \mathcal P [ F\wedge C_4]\Bigg\} =\frac{T_5}{g_s}\int d^6 \zeta \, {\cal L}
\end{equation}
where, as usual,  $\mathcal P$ denotes the pullback of the different spacetime fields on the worldvolume of the brane, $F$ is the field strength of the worldvolume gauge field and $T_5$ is the tension of the D5-brane. 
We will be working we work in units in which the radius of the $AdS_5$ 
is taken to be 1. In such units
the tension of the brane  and the string coupling are given by
\begin{equation} \label{def-tension-coupling}
T_5 = \frac{\lambda^{3/2}}{(2 \, \pi)^5} \quad \& \quad g_s = \frac{g_{YM}^2}{4\, \pi} \quad {\rm with} \quad \lambda = \alpha'^{-2} \, . 
\end{equation}
In \eqref{D5-Eucl} we have also used the fact  that the 
$B$-field vanishes for the $AdS_5\times S^5$ solution of the 
type IIB supergravity equations.

We take the world-volume coordinates of a D5-brane to be $\hat\rho,\theta_1,\phi,\psi,$ $ \beta, \gamma$ and consider the following ansatz for the rest of the coordinates
\begin{equation}\label{embedding}
\tilde\psi =0 \, , \quad 
{\tilde \beta} = \frac{\pi}{2} ,\quad 
{\tilde \gamma} = 0 \quad \& \quad
u= u_0 \, . 
\end{equation}
%
%In figure \ref{figg-1} we depict the D5 brane with the coordinates $x_0, x_1, \beta$ and $\gamma$ suppressed. 
%Let us make a brief comment on  the relation between
%the parameter  $\s$ of our solution and the physical parameters $k$ and $\lambda$. In the limit of small $\s$  the flux $k$ becomes very large and the inclination of the brane with respect to the $AdS_5$ boundary goes to zero. On the other hand, when $\s=1$, which is the endpoint of the range of validity of our solution, the inclination angle becomes $\pi/4$ (see figure   \ref{figg-1}).

The solution above is supported by a worldvolume 2-form flux through the 2-sphere $S^2$ that is parametrised by the angles $(\beta,\gamma)$. The corresponding  1-form gauge field living on the worldvolume of the D5 brane is given by
\begin{equation}\label{A}
A = -\frac{\kappa}{2 \, \pi \, \alpha'} \, \cos \beta \, d\gamma .
\end{equation} 
The functional dependence of the gauge field $A$ is determined by the requirement that its equation of motion  is satisfied.
Furthermore, one can check that, given the ansatz in \eqref{embedding}, the equations of motion derived from the action \eqref{D5-Eucl} 
are automatically satisfied  for the coordinates  $\tilde \beta$ and $\tilde \gamma$. Then, the equation of motion of the coordinate $u$ determines the value of the constant $\kappa$ to be
\begin{equation}\label{kappa}
\kappa=\frac{4+\s^2}{\s \sqrt{8-\s^2}} \, , \quad {\rm where}\quad
\, \sinh u_0\equiv\frac{1}{\s}\, .
\end{equation}
The quantised flux through $S^2$ is
\begin{equation}\label{kappa-1}
k=\int_{S^2} \frac{F}{2 \, \pi} \quad \Rightarrow \quad 
k=\frac{1}{2 \pi}\frac{ \kappa}{2 \pi \alpha'}\int_0^\pi\sin \beta\,d\beta \int_0^{2\pi}d\gamma \quad \Rightarrow \quad k=\frac{\kappa}{\pi \a'}=\frac{\kappa \sqrt{\lambda}}{\pi}\, ,
\end{equation}
with $k \in {\mathbb N}^*$.

Notice that this is precisely the gauge field of the D5 brane solution of \cite{Georgiou:2025mgg}.
To recapitulate we have rewritten the solution of \cite{Georgiou:2025mgg} in Euclidean signature.  In the coordinate system \eqref{metric-Eucl} the defect wraps the $S^2$ parametrised by $(\theta_1,\phi)$ and   sits at the boundary of the $AdS_3\times S^1$ spacetime which is at $\hat\rho \rightarrow \infty$.

\subsection{Anomaly coefficient at strong coupling}\label{strong-1}

We now proceed to the calculation of the anomaly coefficient $b$ at the strong coupling regime.
To this end, we evaluate the on-shell Lagrangian density of the D5 brane which is given by
%\be\label{Lag-on-shell-0}
%{\cal L}_{on-shell}=\frac{(1+\s^2)^2\, (8+\s^2)}{ \sigma ^5\,\sqrt{ 8-\s^2} } \, 
%\sin \beta \sin\theta_1 \sinh^2\hat\rho\, . 
%\ee
\be\label{Lag-on-shell}
{\cal L}_{on-shell}=\frac{\left(-1+2 \s^2\right) \sin \beta \sin\theta_1 \sinh ^2\hat\rho }{\sigma ^3 \sqrt{8-\sigma ^2}}
\ee
Then, the corresponding Euclidean action reads
%\be\label{action-on-shell-0}
%S^{(E)}_{D5}=\int d\hat\rho \,d\psi \,d\theta_1 \,d\phi\, d\beta \,d\gamma \,{\cal L}_{on-shell}= (2\pi)^3
%2^2\,\frac{(1+\s^2)^2\, (8+\s^2)}{ \sigma ^5\,\sqrt{ 8-\s^2} } \, \int_0^{\hat\rho_0}\,d\hat\rho\, \sinh^2\hat\rho.
%\ee
\be\label{action-on-shell}
S^{(E)}_{D5}= \frac{T_5}{g_s}\int d\hat\rho \,d\psi \,d\theta_1 \,d\phi\, d\beta \,d\gamma \,{\cal L}_{on-shell}=  \frac{T_5}{g_s}(2\pi)^2
2\,\frac{\left(-1+2 \s^2\right)  }{\sigma ^3 \sqrt{8-\sigma ^2}}\, vol(AdS_3),
\ee
where $vol(AdS_3)=\int d\hat\rho \, d\theta_1 \,d\phi\  \sinh^2\hat\rho\, \sin\theta_1$.
In \eqref{action-on-shell} the factor of $(2\pi)^2$ originates from the integration of the isometric directions $\psi$ and $\gamma$ while the  factor of $2$ from the integration over the angle $\beta$. 
We now use the fact that the regularised volume of the unit-radius $AdS_3$ space whose boundary is $S^2$ is
\be\label{vol}
vol(AdS_3)=-2 \pi \ln(\L\, a_\S),
\ee
where the cut-off $\L\rightarrow \infty$  and $a_\S$ is the radius
of $S^2$. In the dual gauge theory side $\L$ is the UV cut-off. Putting everything together we obtain
\be\label{action-on-shell-fin}
S^{(E)}_{D5}=-\frac{2}{\pi}\,\frac{\left(-1+2 \s^2\right)  }{\sigma ^3 \sqrt{8-\sigma ^2}}\, \sqrt{\l}\,N\,\ln(\L\, a_\S).
\ee
By taking into account that for a two-sphere $S^2$, $\int_{S^2} d^2\sigma \sqrt{\g}  \, {\cal{R}}_{def} = 8 \pi$ and by combining \eqref{coeffs} and \eqref{Wdiv} we deduce that (see also  \cite{Jiang:2024wzs})
\be\label{Tseyt}
S_E^{(D5)}=-\frac{b}{3}\,\log(\L \, a_\S)\, .
\ee
Direct comparison of \eqref{Tseyt} and \eqref{action-on-shell-fin} gives for the anomaly coefficient $b$ at strong coupling the value 
\be\label{b-1}
b^{(strong)}=\frac{6}{\pi}\,\frac{\left(-1+2 \s^2\right)  }{\sigma ^3 \sqrt{8-\sigma ^2}}\, \sqrt{\l}\,N\, .
\ee
Notice the behaviour of the coefficient $b$ as a function of the parameter of the solution $\s$. For small values of $\s$, namely for 
$\s<{1 \over \sqrt{2}}$, $b$ is negative $b^{(strong)}<0$, at 
$\s={1 \over \sqrt{2}}$ it becomes zero, i.e.  $b^{(strong)}(\s={1 \over \sqrt{2}})=0$ while for $\s>{1 \over \sqrt{2}}$ is positive $b^{(strong)}>0$.
At this point, let us stress that, as mentioned in \cite{Jensen:2018rxu}, unitarity does not require $b\ge 0$. However, in almost all the known cases $b>0$  \cite{Jensen:2018rxu}. To the best of our knowledge, the only exception is that of a {\it free} massless scalar with
Dirichlet boundary conditions, which has $b<0$. The dCFT of \cite{Georgiou:2025mgg} the first example of an {\it interacting} defect conformal field theory with $b<0$. Furthermore, it provides a theory in which the sign of $b$ changes as one of the parameters of the dCFT varies.

Before proceeding let us add some clarifications regarding the regularised $AdS_3$ volume of \eqref{vol}.
 Literally speaking, in the coordinate system we are using the defect is situated at the boundary of the $AdS_3$ which is at $\hat\rho=\infty$. 
Here instead of integrating all the way  to infinity, which makes the action to diverge, we place a large but finite cut-off $\hat\rho_0$  for the radial coordinate $\hat\rho$.
%This cut-off is related to the distance $\epsilon$ of the defect from the boundary 
One may use another coordinate system in which the $AdS_5$ part of the metric \eqref{metric-Eucl} takes the form
\begin{equation}\label{metric-Eucl-1}
ds_{AdS_5}^2 = \frac{1}{z^2}\, \Big[dz^2 +dr^2+ r^2(d\theta_1^2+\sin^2\theta_1 d\phi^2)
+dx^2 \Big] \,.
\end{equation}
The two coordinate systems are related by 
\be\label{coord-rel}
z=\frac{a_\S}{\cosh u\, \cosh\hat\rho-\sinh u\, \cos\psi},\quad r=z \cosh u \sinh\hat\rho, \quad x=z  \sinh u \sin \psi \, .
\ee
In the coordinate system of \eqref{metric-Eucl-1} the defect is situated at the boundary at $z=\epsilon$, at the radius $r=a_\S$ and wraps the $S^2$ parametrised by $(\theta_1,\phi)$.
From the second equation in \eqref{coord-rel} we deduce that the cut-off $\hat\rho_0$ is given by $\sinh \hat\rho_0=\frac{a_\S}{\epsilon \cosh u_0}$ which at the limit $\epsilon \rightarrow 0\Rightarrow \hat\rho_0\rightarrow \infty$ implies $\hat\rho_0=\ln\frac{2 a_\S}{\epsilon \cosh u_0}$.
One can now evaluate the volume of the $AdS_3$ space to be 
\be\label{volads}
vol(AdS_3)=\int d\hat\rho \, d\theta_1 \,d\phi\  \sinh^2\hat\rho\, \sin\theta_1=4 \pi \int_0^{\hat\rho_0} d\hat\rho \sinh^2\hat\rho=-4 \pi \Big(\frac{\hat\rho_0}{2}+\frac{1}{4}\sinh 2\hat\rho_0\Big)
\ee
By plugging in the last equation the value for $\hat\rho_0=\ln\frac{2 a_\S}{\epsilon \cosh u_0}$ one obtains
\be\label{volads-fin}
vol(AdS_3)= -2 \pi \ln\frac{a_\S}{\epsilon}+ 2 \pi\,( \frac{ a_\S^2}{\epsilon^2 \cosh^2 u_0}-\ln\frac{2 }{ \cosh u_0})\, .
\ee
The first term in \eqref{volads-fin} is precisely the relevant for our calculation regularised $AdS_3$ volume of \eqref{vol} while the remaining divergent term can be cancelled by adding appropriate counterterms to the action.

\subsection{Anomaly coefficient at weak coupling}\label{weak-1}
We now turn to the calculation of the anomaly coefficient $b$ at the weak coupling regime.
 When using the metric \eqref{metric-Eucl} the dual field theory $\mathcal{N}=4$  SYM lives in the $AdS_3\times S^1$ boundary of the 10-dimensional background. 
 The metric of the space in which the field theory lives is given by
 \be\label{AdS3xS1}
 ds^2=  d\hat \rho^2+ \sinh^2\hat \rho(d\theta_1^2+\sin^2\theta_1 d\phi^2
+d\psi^2 ) \,.
 \ee
 By taking into account that the scalar fields of $\mathcal{N}=4$  SYM transform as weight one fields under Weyl transformations one can translate the profile of the scalars living in ${\mathbb R}^4$
 that is given in \cite{Georgiou:2025mgg} to that of the scalars living  in $AdS_3\times S^1$
 \begin{eqnarray}\label{vevs-1}
&& \varphi_{i+3}^{\text{cl}}\left(\hat\rho,\psi\right) = \left[\begin{array}{cc} \left(t_i\right)_{k\times k} & 0_{k\times \left(N - k\right)} \\ 0_{\left(N - k\right)\times k} & 0_{\left(N - k\right)\times \left(N - k\right)} \end{array}\right] \nonumber \\
&& \varphi_{i}^{\text{cl}}\left(\hat\rho,\psi\right) = 0,   \quad \quad\quad \quad \quad \quad \quad \quad \quad \quad \quad \quad \quad\quad i=1,2,3.
 \end{eqnarray}
 To proceed we write down the Lagrangian of $\mathcal{N}=4$  SYM living in the Euclidean $AdS_3\times S^1$ space. When the field theory lives on the curved space $AdS_3\times S^1$ one needs to include mass terms for the scalar fields. Consequently, the Lagrangian density takes the form
 \begin{equation} \label{ActionSYME}
\LL_{\N = 4}^{(E)}= \frac{2}{g_{\text{\scalebox{.8}{YM}}}^2} \text{tr}\bigg\{\frac{1}{4} F_{\mu\nu} F^{\mu\nu} +\frac{1}{2} \left(D_{\mu}\varphi_i\right)^2 -
 \frac{1}{4}\left[\varphi_i,\varphi_j\right]^2+\frac{1}{2}\frac{R^{(4)}}{6}\varphi_i^2\bigg\} \, ,
\end{equation}
 where $R^{(4)}=-6$ is the Ricci scalar of $AdS_3\times S^1$ and we have set all the fermions to zero.  The corresponding Euclidean action then becomes
   \begin{equation} \label{ActionSYME}
S_{\N = 4}^{(E)} = \frac{2}{g_{\text{\scalebox{.8}{YM}}}^2}\int d\hat\rho \,d\psi \,d\theta_1 \,d\phi\ \text{tr}\bigg\{
 -\frac{1}{4}(\left[\varphi_i,\varphi_j\right]^2+2 \varphi_i^2)\bigg\} \sin\theta_1 \sinh ^2\hat\rho\, ,
\end{equation}
because the kinetic terms in the action vanish.
Employing now the identities \eqref{comm-1}   listed in appendix \ref{Appendix:details_matrices} we arrive at
 \begin{equation} \label{ActionSYME-fin}
S_{\N = 4}^{(E)} = -2\pi \frac{2}{g_{\text{\scalebox{.8}{YM}}}^2}\frac{k(k^2-1)}{32} vol(AdS_3)=\frac{(2 \pi)^2}{g_{\text{\scalebox{.8}{YM}}}^2}\frac{k(k^2-1)}{16} \ln(\L \, a_\S)\, ,
\end{equation}
 where in the last equality we have use the expression for the regularised $AdS_3$ volume \eqref{vol}.
Then, from \eqref{ActionSYME-fin} and $S_{\N = 4}^{(E)} =-\frac{b}{3}\,\ln(\L \, a_\S)$ one gets for the anomaly coefficient of the intrinsic curvature at weak coupling
\be\label{b-1-weak}
b^{(weak)}=-3 \frac{ \pi^2}{g_{\text{\scalebox{.8}{YM}}}^2}\frac{k(k^2-1)}{4} \, .
\ee
Please notice that, as in the result in the strong coupling regime, $b^{(weak)}< 0$.
%As mentioned in \cite{Jensen:2018rxu} ...

\begin{comment}
\begin{equation}\label{metric-Eucl-1}
ds^2=  dr^2+ r^2(d\theta_1^2+\sin^2\theta_1 d\phi^2)
+dx^2  \,.
\end{equation}

\begin{equation}\label{metric-Eucl-1}
ds^2=  \tilde r^2 (d\hat \rho^2+ \sinh^2\hat \rho(d\theta_1^2+\sin^2\theta_1 d\phi^2)
+d\psi^2 ) \,.
\end{equation}
The two coordinate systems are related by 
\be\label{coord-rel}
\tilde r=\frac{1}{ \cosh\hat\rho-\cos\psi},\quad r=\tilde r\sinh\hat\rho, \quad x=\tilde r \sin \psi \, .
\ee
\end{comment}

\subsection{Comparison and agreement}\label{comp-1}

In this section, we compare the results for the anomaly coefficient $b$ at the weak and strong coupling regime and find that they agree in a certain limit.

The situation here resembles the BMN limit in which one considers operators with large R-charge $J$ and defines the effective coupling $\l'=\frac{\l}{J^2}$ which can be kept 
small in both field and string theory. Then one compares the observables order by order in $\l'$.
In analogy, the small quantity here will be taken to be $\frac{\l}{k^2}$, where $k$ is the units of flux through the two-sphere. Equivalently, one has $\frac{k}{\sqrt{\l}}=\frac{\kappa}{\pi}\gg1$ which by taking into account \eqref{kappa} implies that 
\be\label{kappapprox}
k\approx\frac{\sqrt{2 \l}}{\pi \s}\qquad {\rm when} \qquad\s \rightarrow 0
\ee
since the  condition $\frac{\kappa}{\pi}\gg1$ can be only realised for $\s\rightarrow 0$. We remind the reader that the D5 brane is unstable for $\s>1$ and thus the condition $\frac{\kappa}{\pi}\gg1$ can not be realised for $\s\rightarrow 2 \sqrt{2}$ \cite{Georgiou:2025mgg}.
By plugging \eqref{kappapprox} in \eqref{b-1-weak} and \eqref{b-1} we obtain
\be\label{agreement-1}
b^{(weak)}=b^{(strong)}=-\frac{3}{\sqrt{2}\pi \s^3} \sqrt{\l} N=-3 \frac{ \pi^2}{g_{\text{\scalebox{.8}{YM}}}^2}\frac{k^3}{4} \, .
\ee
We conclude that in the aforementioned limit \eqref{kappapprox} the values of  the anomaly coefficient $b$ at weak and strong coupling agree and that this common value is negative $b<0$. 

\section{Type-A defect anomaly coefficient for the defect CFT of \cite{Georgiou:2025wbg}}\label{anom-sol-1}

\subsection{The D5 brane of \cite{Georgiou:2025wbg} in Euclidean signature}\label{D5-sol-2}
In this section  we rewrite the D5-brane solution of \cite{Georgiou:2025wbg} which realises the gravity duals of certain generically non-supersymmetric 
co-dimension 2 defect conformal field theories in Euclidean signature. 

The solution depends on two independent continuous parameters and ends on a two-dimensional submanifold of the $AdS_5$ boundary.  Consequently, it provides the realisation of the holographic dual of a co-dimension 2 defect conformal field theory (dCFT). 
The D5 probe brane wraps an $S^2$  of the internal space $S^5$. 
The symmetry of the induced, on the brane metric, is 
$AdS_3\times S^1\times S^2$. More details for this solution can be found in \cite{Georgiou:2025wbg}.
The brane orientation of the co-dimension 2 D3-D5 probe-brane system 
%(see \eqref{metric} for the metric and \eqref{embedding} for the embedding ansatz of the D5-brane) 
is given in table \ref{Table:C2D3D5sys}. 
\begin{table}[H]
\begin{center}\begin{tabular}{|c||c|c|c|c|c|c|c|c|c|c|}
\hline
& ${\color{red}\hat\rho}$ & ${\color{red}\theta_1}$ & ${\color{red}\phi}$ & ${\color{red}\psi}$ & ${\color{red}u}$ & ${\color{blue}\tilde \psi}$ & ${\color{blue} {\tilde \beta} }$ & ${\color{blue} {\tilde{ \gamma}}}$ & ${\color{blue}\beta}$ & ${\color{blue}\gamma}$ \\ \hline
\text{D3} & $\bullet$ & $\bullet$ & $\bullet$ & $\bullet$ &&&&&& \\ \hline
\text{D5 probe} & $\bullet$ & $\bullet$ & $\bullet$ &  $\bullet$&   & & &  & $\bullet$ & $\bullet$ \\ \hline
%\text{D7 probe} & $\bullet$ & $\bullet$ & $\bullet$ & &$\bullet$   & &$\bullet$&$\bullet$ & $\bullet$ & $\bullet$ \\ \hline
\end{tabular}
\caption{The D3-D5 intersection. \label{Table:C2D3D5sys}}\end{center}
\end{table}

We take the world-volume coordinates of a D5-brane to be $\zeta^\mu = (\hat\rho,\theta_1,\phi, \psi, \beta, \gamma)$ and consider the following ansatz for the rest of the coordinates
\begin{equation}\label{embedding-2}
\tilde\psi =\tilde\psi_0 \, , \quad 
{\tilde \beta} = \frac{\pi}{2} ,\quad 
 {\tilde \gamma} =\frac{\psi-\phi_0}{\rho} \quad \& \quad
u= u_0 =\sinh^{-1}\frac{1}{\s}\, . 
\end{equation}
%
%In figure \ref{figg-1} we depict the D5 brane with the coordinates $x_0, x_1, \beta$ and $\gamma$ suppressed. 
%Let us make a brief comment on  the relation between
%the parameter  $\s$ of our solution and the physical parameters $k$ and $\lambda$. In the limit of small $\s$  the flux $k$ becomes very large and the inclination of the brane with respect to the $AdS_5$ boundary goes to zero. On the other hand, when $\s=1$, which is the endpoint of the range of validity of our solution, the inclination angle becomes $\pi/4$ (see figure   \ref{figg-1}).

The solution above is supported by a worldvolume 2-form flux through the 2-sphere $S^2$ that is parametrised by the angles $(\beta,\gamma)$. The corresponding  1-form gauge field living on the worldvolume of the D5 brane is given by
\begin{equation}\label{A}
A =- \frac{\kappa}{2 \, \pi \, \alpha'} \, \cos \beta \, d\gamma .
\ee
As in the previous section, the  functional form of $A$ is determined by the requirement that the equation of motion for the potential $A$ is satisfied.
One can check that, given the ansatz in \eqref{embedding-2}, the equations of motion derived from the action \eqref{D5-Eucl} 
are satisfied automatically for the coordinates  $\tilde \beta$ and $\tilde \gamma$. The equation of motion of the coordinate $u$ determines the value of the constant $\kappa$ to be
\begin{equation}\label{kappa-2}
\kappa = \frac{1}{\sigma} \, \cos\tilde \psi_0 
\, \sqrt{\big. \left(2-3 \cos^2\tilde \psi_0 \right) 
\s^2 +2 \rho ^2}\, .
\end{equation}
Finally, the equation of motion for the $S^5$ angle  $\tilde \psi$ forces the constant angle $\tilde \psi_0$ to take the following value
\begin{equation}\label{tildepsi}
\cos^2\tilde \psi_0= \frac{2 \, \sqrt{2}}{ 3 \, \sigma} 
\, \rho \, \sqrt{ \big. \left(\rho ^2-1\right) 
\left(1+\sigma ^2\right)}- (\rho ^2-1)\, .
\end{equation} 
Our solution carries $k \in {\mathbb N}^*$ units of flux through the $S^2$ given by 
\begin{equation}\label{kappa-1}
k=\int_{S^2} \frac{F}{2 \, \pi} \quad  \Rightarrow \quad k=\frac{\kappa}{\pi \a'}=\frac{\kappa \sqrt{\lambda}}{\pi}\, .
\end{equation}

Notice that this is precisely the gauge field of the D5 brane solution of \cite{Georgiou:2025wbg}.
To recapitulate we have rewritten the solution of \cite{Georgiou:2025wbg} in Euclidean signature.  In the coordinate system \eqref{metric-Eucl} the defect wraps the $S^2$ parametrised by $(\theta_1,\phi)$ and   sits at the boundary of the $AdS_3\times S^1$ spacetime which is at $\hat\rho \rightarrow \infty$. 

\subsection{Anomaly coefficient at strong coupling}\label{strong-2}

In this section we present the calculation of the $b$ anomaly coefficient of the D5 brane presented in \cite{Georgiou:2025wbg} whose Euclidean version is described in section \ref{D5-sol-2}.

We start by evaluating the on-shell Lagrangian density of this solution which reads
\begin{eqnarray}\label{Lag-on-shell}
&&{\cal L}_{on-shell}=\frac{1}{\rho  \sigma ^5}  \sin\beta   \sin \theta_1 \sinh ^2\hat\rho  \left(\sigma ^2+1\right)^{3/2}\\
&&\Bigg( -\rho  \sigma  \frac{\left(2 \sigma ^2+1\right)}{ \left(\sigma ^2+1\right)^{3/2}} \kappa+\sqrt{2} \rho^2 \cos \tilde\psi_0 +\sqrt{2} \sigma ^2 
\cos \tilde\psi_0 \sin ^2\tilde\psi_0\Bigg).  \nonumber
\end{eqnarray}

In a similar fashion to that of section \ref{strong-1}, the on-shell action for the Euclidean version of the D5 brane of \cite{Georgiou:2025wbg} becomes
\begin{eqnarray}\label{action-on-shell}
&&S^{(E)}_{D5}= \frac{T_5}{g_s}\int d\hat\rho \,d\psi \,d\theta_1 \,d\phi\, d\beta \,d\gamma \,{\cal L}_{on-shell}=  \frac{T_5}{g_s}(2\pi)^2
2\,\frac{   \left(\sigma ^2+1\right)^{3/2}}{\rho  \sigma ^5}   \nonumber\\
&&
\Bigg( -\rho  \sigma  \frac{\left(2 \sigma ^2+1\right)}{ \left(\sigma ^2+1\right)^{3/2}} \kappa+\sqrt{2} \rho^2 \cos \tilde\psi_0  +\sqrt{2} \sigma ^2 
\cos \tilde\psi_0  \sin ^2\tilde\psi_0\Bigg)\, vol(AdS_3), 
\end{eqnarray}
From the last equation one can read the anomaly coefficient $b$ after inserting the expressions for the regularise $AdS_3$ volume \eqref{vol}, as well as the one for the brane tension \eqref{def-tension-coupling}. It reads
\be\label{b-2}
b^{(strong)}=\frac{6}{\pi}\,\frac{   \left(\sigma ^2+1\right)^{3/2}}{\rho  \sigma ^5} \Bigg( -\rho  \sigma  \frac{\left(2 \sigma ^2+1\right)}{ \left(\sigma ^2+1\right)^{3/2}} \kappa+\sqrt{2} \rho^2 \cos \tilde\psi_0  +\sqrt{2} \sigma ^2 
\cos \tilde\psi_0  \sin ^2\tilde\psi_0\Bigg)\, \sqrt{\l}\,N\, .
\ee

Let us now comment on this expression for the coefficient $b$. As discussed in \cite{Georgiou:2025wbg} the  D5 brane solution introduced there interpolates between two special dCFTs. At one end $\cos\tilde\psi_0=1$, the D5 brane solution of \cite{Georgiou:2025wbg} becomes that of \cite{Georgiou:2025mgg}. In agreement with that observation the coefficient  of \eqref{b-2} should become that of \eqref{b-1}. By taking into account that in the $\cos\tilde\psi_0=1$ limit, $\rho$ and $\kappa$ become equal to $\rho=\sqrt{\frac{8 \sigma ^2+8}{8-\sigma ^2}}$ and $\kappa=\frac{4+\s^2}{\s \sqrt{8-\s^2}}$ (see section 2.2 of  \cite{Georgiou:2025wbg}), it is easy to verify that this is, indeed, the case. At the other end $\cos\tilde\psi_0=0$ it is straightforward to see that $b^{(strong)}(\tilde\psi_0=\pi/2)=0$ because in this limit $\kappa=0$.
In section \ref{weak-2}, we will see that the weak coupling result for $b$ exhibits a similar behaviour.

\subsection{Anomaly coefficient at weak coupling}\label{weak-2}
The classical solution to the $\mathcal{N}=4$ SYM equations of motion which is dual to the D5 brane of the last section is given by \cite{Georgiou:2025wbg}
\begin{eqnarray}\label{sol-2}
&&\varphi_{i+3}^{\text{cl}}\left(\hat\rho,\psi\right) =    \left[\begin{array}{cc} \left(t_i\right)_{k\times k} & 0_{k\times \left(N - k\right)} \\ 0_{\left(N - k\right)\times k} & 0_{\left(N - k\right)\times \left(N - k\right)} \end{array}\right], \quad i=1,2,3 
\\
&&\varphi_{1}^{\text{cl}}\left(\hat\rho,\psi\right) = 0,\quad \varphi_{2}^{\text{cl}}(\hat\rho,\psi) +i \varphi_{3}^{\text{cl}} (\hat\rho,\psi)=   \sin{\tilde \psi}_0 \,e^{i \psi}\,C\, , \nonumber 
\end{eqnarray}
with the diagonal constant matrix $C$ being
\be\label{C-matr}
C=\left(\begin{array}{cc} (\b_1+i \g_1) \otimes I_{(N_1+k)\times (N_1+k)} &\,\,\,\, 0_{(N_1+k)\times \left(N-N_1 - k\right)} \\ 0_{\left(N-N_1- k\right)\times (N_1+k)} &\,\,\,\,  0_{\left(N-N_1 - k\right)\times \left(N-N_1 - k\right)} \end{array}\right)_{N\times N}\, .  
\ee
As in section \ref{weak-1}  the field theory is taken to reside on $AdS_3\times S^1$ with metric \eqref{AdS3xS1}.
For details regarding the field theory dual of the D5 brane of section \ref{strong-2} we refer the reader to \cite{Georgiou:2025wbg}.
Here let us only mention that $N_1$ is the number of the D5 probe branes.

In order to derive the on-shell action one should substitute the solution \eqref{sol-2} in the general expression for the Lagrangian density \eqref{ActionSYME}.
In doing so, one gets two separate contributions. The first originates from the three scalars $\varphi_{i},\, i=1,2,3$ while the second from the  remaining three  $\varphi_{i+3},\, i=1,2,3$. 
It is straightforward to show the first contribution is zero. This is so because 
\be
\sum_{i=1}^3\left(D_{\mu}\varphi_i\right)^2=\sum_{i=1}^3\left(\partial_{\mu}\varphi_i\right)^2=\sum_{i=1}^3\varphi_i^2\, .
\ee
The last equation together with the fact that the commutator involving any of the first three scalar fields is zero nullifies their contribution to the action. \footnote{Notice also that the commutator 
$\left[\varphi_{i+3},\varphi_j\right]=0$ for $i=1,2,3$ and  $j=1,2,3$. This is the reason why the two contributions, one from the scalars $\varphi_i,\, i=1,2,3$ and the other from $\varphi_i,\, i=4,5,6$, disentangle.}

The calculation of the remaining non-zero contribution is the same as that of section \ref{weak-1}. The  final result is 
\begin{equation} \label{ActionSYME-fin-2}
S_{\N = 4}^{(E)} = -2\pi \frac{2}{g_{\text{\scalebox{.8}{YM}}}^2}\frac{k(k^2-1)}{32} vol(AdS_3)=\frac{(2 \pi)^2}{g_{\text{\scalebox{.8}{YM}}}^2}\frac{k(k^2-1)}{16} \ln(\L \,a_\S)\, ,
\end{equation}
but with $k=\frac{\kappa \sqrt{\l}}{\pi}$, where now $\kappa$ given by \eqref{kappa-2}.
From \eqref{ActionSYME-fin-2} and $S_{\N = 4}^{(E)} =-\frac{b}{3}\,\ln(\L \, a_\S)$ we derive the anomaly coefficient $b$ at weak coupling to be
\be\label{b-2-weak}
b^{(weak)}=-3 \frac{ \pi^2}{g_{\text{\scalebox{.8}{YM}}}^2}\frac{k(k^2-1)}{4} \, .
\ee
Notice that the weak coupling result is functionally the same as that of the dCFT of \cite{Georgiou:2025mgg}, see \eqref{b-1-weak}, because the contribution from the three first scalars $\varphi_i,\, i=1,2,3$ is zero. However, in each case $k$ depends differently on the parameters of the dual D5 brane solution. For the case at hand $k$ is given by \eqref{kappa-1} and \eqref{kappa-2} while for the case \cite{Georgiou:2025mgg} - equation \eqref{b-1-weak} - by \eqref{kappa}.

The behaviour of \eqref{b-2-weak} is in accordance with the interpolating structure of the corresponding dCFT. At the end where $\cos\tilde\psi_0=0$, the coefficient $b$ vanishes, i.e. $b^{(weak)}(\tilde\psi_0=\pi/2)=0$, as it happens for the strong coupling result $b^{(strong)}(\tilde\psi_0=\pi/2)=0$. This vanishing is due to the corresponding vanishing of $k$ (see discussion above equation (5.19)  of \cite{Georgiou:2025wbg}). At the other end in which  $\cos\tilde\psi_0=1$ the weak coupling result becomes exactly the same to that of \eqref{b-1-weak} in agreement with the fact that at this limit 
the dCFT of \cite{Georgiou:2025wbg} becomes that of \cite{Georgiou:2025mgg}.

\subsection{Comparison and agreement}\label{comp-2}
In this section, we compare the expressions for the anomaly coefficient $b$ at strong and weak coupling which were derived  in the last two sections. The comparison and agreement 
will be  in the limit \eqref{kappapprox} discussed in section \ref{comp-1}. However, in this case  implementing the above limit is a bit intricate because 
the condition $\s\rightarrow 0$ enforces $\rho\rightarrow 1$. 
This can be seen from the fact that
the requirement  $0\leq \cos{\psi_0}\leq 1$ implies the following inequality between  $\rho$ and $\s$
 \begin{eqnarray}\label{boundss}
1 \leq \rho \leq \sqrt{\frac{8 \sigma ^2+8}{8-\sigma ^2}}\qquad {\rm when} \qquad 0<\sigma <2 \, \sqrt{2}\, .
\end{eqnarray}
So when sending $\s\rightarrow 0$ \eqref{boundss} implies that $\rho\leq1+\frac{9}{16} \s^2+...$.
The comparison now is done around the solution having $\cos{\psi_0}=1$. 
To this end we set 
\be\label{ro}
\rho=\sqrt{\frac{8 \sigma ^2+8}{8-\sigma ^2}}-\d,\, \quad\d\ll1.
\ee
 As a result we have the refined double scaling limit, namely 
\be\label{dslimit}
\d\rightarrow 0,\quad  \s\rightarrow 0 \qquad{\rm with} \qquad {\rm fixed}=\bar\epsilon={\d \over \s^2}\ll1
\ee
When \eqref{ro} is inserted in \eqref{tildepsi} we obtain 
\begin{eqnarray}\label{cos}
&&\cos{\psi_0}=1-\frac{ \d  \sqrt{2(\sigma ^2+1)(8-\sigma ^2)}}{9 \sigma ^2}-\frac{\d ^2 \left(\sigma ^4-10 \sigma ^2+16\right)}{54 \sigma ^4}+
{\mathcal O}(\bar\epsilon^3)\simeq\nonumber \\
&&1-\frac{4 }{9 }\bar\epsilon-\frac{8 }{27}\bar\epsilon^2+{\mathcal O}(\bar\epsilon)^3).
\end{eqnarray}
As we mentioned above, to obtain the last equality we have taken the limit $\s \rightarrow 0$ by keeping $\bar\epsilon={\d \over \s^2}$ fixed and much less than the unity.

By performing the analogous expansions for all the quantities  up to the desired order one obtains from \eqref{b-2}
\begin{eqnarray}\label{b-3}
b^{(strong)}={\sqrt{\l}\,N \over \pi  \sigma ^3}&&\Big(-\frac{3}{\sqrt{2} } +  2 \sqrt{2}\, \bar\varepsilon+ \frac{4 \sqrt{2}}{9 }\, \bar\varepsilon ^2
+\frac{80 \sqrt{2}}{243 }\, \bar\varepsilon ^3+\frac{80 \sqrt{2}}{243 }\, \bar\varepsilon ^4+\frac{832 \sqrt{2}}{2187 }\, \bar\varepsilon ^5
\nonumber\\
&&+\frac{28288 \sqrt{2}}{59049 }\bar\varepsilon ^6+\frac{113152 \sqrt{2}}{177147 }\bar\varepsilon ^7+\frac{1414400 \sqrt{2}}{1594323 }\bar\varepsilon ^8+{\mathcal O}(\bar\epsilon)^9)\Big)\, .
\end{eqnarray} 
{\it Precisely} the same expansion is obtained from manipulating in a similar way the weak coupling $b^{(weak)}$ anomaly coefficient of \eqref{b-2-weak}. Needless to mention that the two expansions differ by terms of order ${1 \over \s^3}\bar\varepsilon^n \s^m$ with $m,n>1$ which, however, vanish in the limit \eqref{dslimit}.
 
 Actually, one can re-sum the series above to obtain the result in a closed form. It reads
 \be\label{agree}
 b^{(strong)}=b^{(weak)}=-{\sqrt{\l}\,N \over \pi  \sigma ^3} \frac{(9-16\, \bar\varepsilon )^{3/4}}{\sqrt{6} }\,. 
 \ee
As in the case of the dCFT of  \cite{Georgiou:2025mgg} the anomaly coefficient $b$ of \cite{Georgiou:2025wbg} is negative in the limit \eqref{dslimit}.

\begin{comment}
By  expanding around $\s=0$ the expression \eqref{b-2-weak} for $b$ at weak coupling one gets
\be\label{b-3-weak}
b^{(weak)}=- \frac{ 2^{7/4}}{\pi}\frac{\rho^{9/2}(\rho^2-1)^{3/4}}{\sqrt{3}\,\s^{9/2}}  \sqrt{\l}\,N\, .
\ee
Similarly by expanding around $\s=0$ the expression \eqref{b-2} for $b$ at strong coupling one gets
\be\label{b-3}
b^{(strong)}=- \frac{ 2^{7/4}}{\pi}\frac{\rho^{1/2}(\rho^2-1)^{3/4}}{\sqrt{3}\,\s^{9/2}}  \sqrt{\l}\,N\, .
\ee
The two expressions seemingly disagree because powers of $\rho$ are different. However, one should not forget that the condition $\s\rightarrow 0$ enforces $\rho\rightarrow 1$. 
This can be seen from the fact that
the condition $0\leq \cos{\psi_0}\leq 1$ implies the following inequality between  $\rho$ and $\s$
 %
 \begin{eqnarray}\label{boundss}
1 \leq \rho \leq \sqrt{\frac{8 \sigma ^2+8}{8-\sigma ^2}}\qquad {\rm when} \qquad 0<\sigma <2 \, \sqrt{2}\, .
\end{eqnarray}
So when sending $\s\rightarrow 0$ \eqref{boundss} implies that $\rho\leq1+\frac{9}{16} \s^2+...$.
So by putting this value for 
$\rho$ the leading terms of \eqref{b-3-weak} and \eqref{b-3} agree and give
\be\label{b-fin1}
b^{(weak)}=b^{(strong)}=- \frac{ 2^{7/4}}{\pi}\frac{(\rho^2-1)^{3/4}}{\sqrt{3}\,\s^{9/2}}  \sqrt{\l}\,N\leq-\frac{3}{\sqrt{2}\,\pi\, \s^3} \sqrt{\l}\,N\, .
\ee
\end{comment}

\section{Type-B defect anomaly coefficient for the defect CFT of \cite{Georgiou:2025wbg} and \cite{Georgiou:2025mgg}}\label{anom-sol-d1}
In this section, we will calculate the type-B anomaly coefficient $d_1$ multiplying the external curvature term of the Weyl anomaly for the dCFT of \cite{Georgiou:2025wbg}, first at strong and then at weak coupling. At the appropriate limit agreement between the weak and strong coupling results is observed. Before embarking on the calculation, let us mention that the anomaly coefficient $d_1$ for the dCFT of  \cite{Georgiou:2025mgg} can be obtained from the analogous expressions for the dCFT of \cite{Georgiou:2025wbg} by setting $\tilde\psi_0=0$ since, as mentioned in \cite{Georgiou:2025wbg}, the  dCFT of \cite{Georgiou:2025mgg} is one of the end points of the more general construction of \cite{Georgiou:2025wbg}.

\subsection{Anomaly coefficient at strong coupling}\label{strong-3}
In this section, we present the calculation at strong coupling. To this end, we employ the Euclidean version of our solution but for the case in which the defect is supported on ${\mathbb R}^2$ instead of $S^2$. However, the extrinsic curvature of ${\mathbb R}^2$ is zero. In order to extract $d_1$ one needs a defect whose geometry supports non-zero extrinsic curvature. To this end, we make small deformations of the defect which sits at the boundary. We then find the form of D5 brane solution close to the boundary.  In other words, we holographically determine the response of the defect to deformations along the perpendicular directions by changing slightly the shape of the defect on the boundary and by subsequently calculating how this deformation propagates into the bulk. It is adequate to find the form of the solution not everywhere but only close to the boundary since it is this part of the solution which produces the divergent logarithmic term in the action.

We start by writing the metric of the Euclidean $AdS_5$ space as 
\be\label{flat-metric}
ds^2=\frac{1}{z^2}(dx_0^2+dx_1^2+dx_2^2+dx_3^2+dz^2)\, .
\ee
As world-volume coordinates of the $D5$ brane we choose $(x_0, x_1, z, \tilde \g, \beta, \gamma)$.
\begin{table}[H]
\begin{center}\begin{tabular}{|c||c|c|c|c|c|c|c|c|c|c|}
\hline
& ${\color{red}x_0}$ & ${\color{red}x_1}$ & ${\color{red}x_2}$ & ${\color{red}x_3}$ & ${\color{red}z}$ & ${\color{blue}\tilde \psi}$ & ${\color{blue} {\tilde \beta} }$ & ${\color{blue} {\tilde{ \gamma}}}$ & ${\color{blue}\beta}$ & ${\color{blue}\gamma}$ \\ \hline
\text{D3} & $\bullet$ & $\bullet$ & $\bullet$ & $\bullet$ &&&&&& \\ \hline
\text{D5 probe} & $\bullet$ & $\bullet$ & &  &  $\bullet$ & & &$\bullet$ & $\bullet$ & $\bullet$ \\ \hline
%\text{D7 probe} & $\bullet$ & $\bullet$ & $\bullet$ & &$\bullet$   & &$\bullet$&$\bullet$ & $\bullet$ & $\bullet$ \\ \hline
\end{tabular}
\caption{The D3-D5 intersection. \label{Table:C2D3D5system}}\end{center}
\end{table}
Then the solution is written as 
\begin{equation}\label{embedding-3}
\tilde\psi =\tilde\psi_0 \, , \quad 
{\tilde \beta} = \frac{\pi}{2} ,\quad 
 x_2 = \frac{1}{\s}\, z \cos(\rho \tilde\g)\quad \& \quad
 x_3 = \frac{1}{\s}\, z \sin(\rho \tilde\g)\, . 
\end{equation}
Furthermore, the value of the gauge field is given by \eqref{A} while the relations between the constants are the same as in \eqref{kappa-2} and \eqref{tildepsi}.
We now make a small deformation of the flat defect given by
\be\label{deform}
x_2 = \frac{1}{\s}\, z \cos(\rho \tilde\g)+\d x_2(x_0, x_1, z, \tilde \g, \beta, \gamma), \quad  \quad  x_3 = \frac{1}{\s}\, z \sin(\rho \tilde\g)+\d x_3(x_0, x_1, z, \tilde \g, \beta, \gamma). 
\ee
The next step is to plug the last equation into the action of the D5 brane and keep the terms which are quadratic in the fluctuations.\footnote{The terms linear in the fluctuations will vanish due to the fact that we are expanding around a solution of the equations of motion.} From this action one can find the equations that govern the evolution of the fluctuations $\d x_2(x_0, x_1, z)$ and $\d x_3(x_0, x_1, z)$, which now are taken to depend only on $x_0$, $x_1$ and $z$.
These equations take the following form
\be\label{flucts}
3 \sigma ^2 \partial_z \d x_i-\sigma ^2 z \,\partial^2_z \d x_i-\left(\sigma ^2+1\right) z \left(\partial^2_{x_0} \d x_i+\partial^2_{x_1} \d x_i\right)=0, \,\,\, \qquad i=2,3 .
\ee
Passing to momentum space $\d x_i(x_0,x_1,z)= \d x_i(z,\ell) e^{i \ell_0 x_0+i \ell_1 x_1}$ converts the last equation to 
\be\label{flucts}
3 \sigma ^2 \partial_z \d x_i(z,\ell)-\sigma ^2 z \,\partial^2_z \d x_i(z,\ell)l+\left(\sigma ^2+1\right) z\,  \ell^2 \d x_i(z,\ell)=0, \,\,\, \qquad i=2,3 .
\ee
The most generic solution of this equation is given by
\be\label{gen-sol}
\d x_i(z,\ell)=c_1 z^2 J_2\left(i \,\ell \, z \sqrt{1+\frac{1}{\sigma ^2}}\right)+c_2 z^2 Y_2\left(-i \,\ell \,z \sqrt{1+\frac{1}{\sigma ^2}}\right), \quad \ell^2=\ell_0^2+\ell_1^2,
\ee
where $J$ and $Y$ are the Bessel functions of the first and second kind, respectively.
The behaviour of the two solutions near the boundary of the space $z=0$ is as follows:
\begin{eqnarray}\label{expan}
&z^2 Y_2\left(-i \,\ell \,z \sqrt{1+\frac{1}{\sigma ^2}}\right)=\frac{4  \sigma ^2}{\pi  \ell^2 \left(\sigma ^2+1\right)}-\frac{ z^2}{\pi }+
{\mathcal O}\left(z^4\right), \nonumber \\
&z^2 J_2\left(i \,\ell \, z \sqrt{1+\frac{1}{\sigma ^2}}\right)= -\frac{1}{8} z^4 \left( \ell^2 \left(\frac{1}{\sigma ^2}+1\right)\right)+{\mathcal O}\left(z^6\right)
\end{eqnarray}
The solution which is relevant for our purposes is the Bessel of the second kind $Y$. This is so because we need a solution which does not vanish when $z=0$ since our aim is to find how a perturbation of the shape of the defect $\d x_{2,3}(x_0,x_1,z=0)=\d x_{2,3}(x_0,x_1)$ propagates in the bulk $z\neq 0$. The propagation is governed by the differential equation \eqref{flucts}. At this point let us stress that we need the behaviour of the solution only close to the boundary since it is this region which will give the logarithmic divergence which is necessary for calculating the anomaly coefficient $d_1$.

Going back to configuration space the relevant to us solution becomes
\be\label{sol-diff}
\d x_i(x_0,x_1,z)= \d x_i(x_0,x_1)+\frac{\left(\sigma ^2+1\right) }{4 \sigma ^2}\, z^2\big(\partial^2_{x_0}\d x_i(x_0,x_1)+\partial^2_{x_2}\d x_i(x_0,x_1)\big)+{\mathcal O}(z^4)\, .
\ee
By comparing to the asymptotic solution of a scalar field in $AdS_{d+1}$, that reads $\phi(x,z)=z^{d-\D}\big(\phi_0(x)+z^2 \phi_2(x)\big)+\cdots$, we see that from the point of view of the D5 brane whose induced metric is $AdS_3$ the fluctuations $\d x_i$ behave as scalars of dimension $\D=2$.

It is now straightforward to substitute the solution \eqref{sol-diff} in to the quadratic Lagrangian to obtain
\begin{eqnarray}\label{Lag-d1}
&S^{(2)}=\frac{T_5}{g_s}\int dx_0 dx_1 dz d\tilde\g d\b d\g \, {\cal L}_{on-shell}^{(2)}= \\
&\frac{T_5}{g_s} \frac{4 \pi^2}{16 \rho  \sigma ^2}\int dx_0 dx_1 \frac{dz}{z} \,2 \,\sqrt{\frac{\left(\sigma ^2+1\right) 
\left(\cos ^4\tilde\psi_0+\kappa ^2\right)}{\rho^2+\s^2 \sin^2 \tilde\psi_0}}\,\Big(\left(2 \sigma ^2+1\right) \sin^2 \tilde\psi_0+\rho ^2\Big)J(x_0,x_1) ,\nonumber
\end{eqnarray}
where
\begin{eqnarray}\label{J}
J(x_0,x_1)=\sum_{i=2,3}\Bigg(\Big(\partial^2_{x_1}\d x_i\Big)^2+2 \,\partial_{x_1}\d x_i \cdot \partial^3_{x_1}\d x_i+2 \,\partial_{x_0}\d x_i \cdot \partial_{x_0}\partial^2_{x_1}\d x_i+\nonumber \\
2 \,\partial^2_{x_1}\d x_i \cdot \partial^2_{x_0}\d x_i+
\Big(\partial^2_{x_0}\d x_i\Big)^2+2 \,\partial_{x_1}\d x_i \cdot \partial_{x_1}\partial^2_{x_0}\d x_i+2 \,\partial_{x_0}\d x_i \cdot \partial^3_{x_0}\d x_i\Bigg)\, .
\end{eqnarray}
In the second line of \eqref{Lag-d1} we have kept only the term which is proportional to $\frac{1}{z}$ because this is the term which upon integration will give the logarithmic term which we are after, namely $\int_\epsilon\frac{dz}{z}=-\ln\epsilon$. Furthermore, the factor of $4 \pi$ originates from the integration over $\b$ and $\g$. Finally, to get the second line of \eqref{Lag-d1} we have also performed the integral over 
$\tilde\g \in [0, \frac{2 \pi}{\rho}]$. The range of integration is chosen so that the boundary angle $\psi \in [0, 2 \pi]$.

One may now observe that after some partial integrations the expression for $J$ becomes proportional to the $Y^{\mu}_{ab}Y_{\mu}^{ab}$.
Indeed, in Euclidean signature and keeping terms quadratic in the small fluctuations $\d x_i, \, i=2,3$, one gets from \eqref{Y}
\begin{eqnarray}\label{Ysq}
&&Y^{\mu}_{ab}Y_{\mu}^{ab}=\sum_{i=2,3}(\partial_a\partial_b \d x_i-\frac{1}{2}\d_{ab}\partial_c\partial^c\d x_i)^2=\nonumber\\ 
&&\frac{1}{2}\sum_{i=2,3}\Big(\Big(\partial^2_{x_0}\d x_i\Big)^2+\Big(\partial^2_{x_1}\d x_i\Big)^2+2 \,\partial^2_{x_1}\d x_i \cdot \partial^2_{x_0}\d x_i \Big)=-\frac{1}{2} \,J(x_0,x_1).
\end{eqnarray}
Consequently,  we see that the effective action of the D5 brane \eqref{Lag-d1} takes the form (see \eqref{Wdiv} and  \eqref{coeffs} )
\begin{eqnarray}\label{action-d1s}
S=\ln\epsilon \int dx_0 dx_1 \, \frac{ d_1}{24 \pi}\,Y^{\mu}_{ab}Y_{\mu}^{ab}+\cdots\, ,
\end{eqnarray}
with the anomaly coefficient $d_1$ being
\be\label{d1-strong-fin}
d_1^{(strong)}=\frac{3 \sqrt{\l }N}{ \pi}\frac{1}{ \rho  \sigma ^2} \,\sqrt{\frac{\left(\sigma ^2+1\right) 
 \left(\cos ^4\tilde\psi_0+\kappa ^2\right)}{\rho^2+\s^2 \sin^2 \tilde\psi_0}}\,\Big((2 \sigma ^2+1) \sin^2 \tilde\psi_0+\rho ^2\Big)>0\,.
\ee
From the last equation we can also obtain the $d_1$ anomaly coefficient at strong coupling for the dCFT of \cite{Georgiou:2025mgg}. By setting $\cos\tilde\psi_0=1$ in \eqref{d1-strong-fin} and by taking into account that in this limit $\rho$ and $\kappa$ become equal to $\rho=\sqrt{\frac{8 \sigma ^2+8}{8-\sigma ^2}}$ and $\kappa=\frac{4+\s^2}{\s \sqrt{8-\s^2}}$ (see section 2.2 of  \cite{Georgiou:2025wbg}) we obtain the following value for the $d_1$ of \cite{Georgiou:2025mgg}
\be\label{d1-of1}
d_1^{(strong)}=\frac{12 \left(\sigma ^2+1\right)}{\pi  \sigma ^3 \sqrt{8-\sigma ^2}}\, .
\ee
At the other end $\cos\tilde\psi_0=0$ it is straightforward to see that $d_1^{(strong)}(\tilde\psi_0=\pi/2)=0$ because in this limit $\kappa=0$. This is due to the fact that in this limit the D5 brane solution becomes singular since the two-sphere which the D5 brane wraps shrinks to zero size.

\subsection{Anomaly coefficient at weak coupling}\label{weak-3}
In this section, we present the calculation of the leading, in the weal coupling expansion, term of the anomaly coefficient $d_1$. To this end, we consider the case in which the defect is a cylinder of radius $a$ embedded in the 4-dimensional  flat Euclidean space with metric
\be\label{flat-metric-FT}
ds^2=(dx_0^2+dr^2+r^2 d\psi^2+dx^2)\, .
\ee
The locus of the defect is at $r=a,\, x=0$ and it is parametrised by the coordinates $(x_0,\phi)$.
We choose the cylinder as the worldsheet of the defect because in contradistinction to the case of the plane $\mathbb R^2$ and the sphere $S^2$ the former has non-zero extrinsic curvature. The main obstacle is that the profile of the scalar fields is known only when the defect is planar (see \cite{Georgiou:2025wbg}). Our strategy will be to find the profile of the scalar fields for the case of the cylinder, not exactly, but as a series expansion in powers of ${1\over a}$ for very large radius of the cylinder $a$. This is adequate and convenient because for $a\rightarrow \infty$ the zeroth order solution is known  since  in this case the defect becomes essentially planar. Then we perturbatively calculate ${1\over a}$ corrections to the zeroth order result.

In order to evaluate the anomaly coefficient $d_1$ we need to solve the following non-linear partial differential equation for the case of the cylinder:
 \be\label{radial}
\frac{\partial^2\varphi_i}{\partial r^2}+\frac{1}{r}\frac{\partial \varphi_i}{\partial r}+\frac{1}{r^2}\frac{\partial^2 \varphi_i}{\partial \psi^2}+\frac{\partial^2\varphi_i}{\partial x^2}= \left[\varphi_j,\left[\varphi_j,\varphi_i\right]\right].
\ee

First we focus on the contribution of the three scalar fields $\varphi_{i+3},\, i=1,2,3$. Later on we shall consider the contribution of the remaining three scalars $\varphi_{i+3},\, i=1,2,3$ (see equation \eqref{sol-2} and \cite{Georgiou:2025wbg}).
By allowing $\varphi_{i+3} \sim t_i, \, i=1,2,3$, as in \eqref{sol-2} and by remembering that the solution for the 3 aforementioned scalars should not depend on the angle $\psi$ due to the cylindrical symmetry of the problem we arrive, after suppressing the index $i$, at the following equation which holds for each one of the scalars
$\varphi_{i+3} \sim t_i, \, i=1,2,3$
 \be\label{radial-1}
\frac{\partial^2\varphi}{\partial r^2}+\frac{1}{r}\frac{\partial \varphi}{\partial r}+\frac{\partial^2\varphi}{\partial x^2}= \varphi^3\, .
\ee
In order to solve iteratively the last equation we will use the following trick. 
First we consider a cylinder with very large radius $a$. Then we solve the equation of motion \eqref{radial-1} order by order in the large $a$ expansion. We will only need the solution near the locus of the defect at $r=a,\, x=0$ since it is the behaviour of the solution near this region which produces the logarithmic term in the action.
The expansion of each of the scalar fields for large $a$ and near the locus of the cylindrical defect reads
\be\label{phi-exp}
\varphi= \varphi_0(\rho,\theta)+\frac{1}{a}\varphi_1(\rho,\theta)+\frac{1}{a^2}\varphi_2(\rho,\theta),
\ee
where we have made the following change of variables from  $(r,x)$ to $(\rho,\theta)$
 \be\label{change-vars}
y=r-a, \,\,\,\, \rho=\sqrt{x^2+y^2}, \, \,\,\, \theta=\tan^{-1}\frac{x}{y}, \,\,\,\, y=\rho \cos\theta, \,\,\,\, x=\rho \sin\theta.
\ee
Accordingly, the expansion of the differential equation \eqref{radial-1} gives\\
$\bullet$ \underline{At order $a^0$}:\\
\be\label{a0}
\frac{\partial^2\varphi_0}{\partial r^2}+\frac{\partial^2\varphi_0}{\partial x^2}= \varphi_0^3 \Longrightarrow \varphi_0=\frac{1}{\rho}=\frac{1}{\sqrt{(r-a)^2+x^2}}.
\ee
$\bullet$ \underline{At order $a^{-1}$}:\\
\begin{eqnarray}\label{a1}
&&\frac{\partial^2\varphi_1}{\partial \rho^2}+\frac{1}{\rho}\frac{\partial \varphi_1}{\partial \rho}+\frac{1}{\rho^2}\frac{\partial^2 \varphi_1}{\partial \theta^2}= \frac{\partial \varphi_0}{\partial r} +3\varphi_1\varphi_0^2=\frac{\cos\theta}{\rho^2}+3\frac{\varphi_1}{\rho^2}
\Longrightarrow \\
&&\varphi_1(\rho,\theta)=(-\frac{1}{4}+c \,\rho^2 )\cos\theta\overset{c=0}{=}-\frac{1}{4}\cos\theta=-\frac{1}{4}\frac{(r-a)}{\sqrt{(r-a)^2+x^2}}.
\end{eqnarray}
The solution above is obtained after performing separation of variables by setting $\varphi_1=A(\rho)\cos\theta$. This separation of variables leads to 
an ordinary differential equation for $A(\rho)$, which reads
\be\label{Ar}
\rho^2 A'' (\rho)+\rho A' (\rho)-4 A (\rho)=1
\ee
The first term in the solution comes precisely from the particular solution of this ordinary differential equation for  $A(\rho)$. The second term, that proportional to $c$, solves the homogeneous equation. In what follows we will set the constant $c=0$ ignoring this second term because such a term does not contribute to the logarithmic term of the action.\\
$\bullet$ \underline{At order $a^{-2}$}:\\
\begin{eqnarray}\label{a2}
&&\frac{\partial^2\varphi_2}{\partial \rho^2}+\frac{1}{\rho}\frac{\partial \varphi_2}{\partial \rho}+\frac{1}{\rho^2}\frac{\partial^2 \varphi_2}{\partial \theta^2}-(r-a)\frac{\partial \varphi_0}{\partial r}+\frac{\partial \varphi_1}{\partial r}= 3\varphi_2\varphi_0^2+ 3\varphi_1^2\varphi_0
\Longrightarrow \nonumber \\
&&\frac{\partial^2\varphi_2}{\partial \rho^2}+\frac{1}{\rho}\frac{\partial \varphi_2}{\partial \rho}+\frac{1}{\rho^2}\frac{\partial^2 \varphi_2}{\partial \theta^2}-3 \frac{ \varphi_2}{\rho^2}=-\frac{1}{32 \rho}(9+17\cos2\theta).
\end{eqnarray}
To solve the last equation we employ separation of variables and write the solution as 
\be\label{ansatz}
\varphi_2(\rho,\theta)=B_1(\rho)+B_2(\rho)\cos2\theta.
\ee
This leads to the following two 
ordinary differential equations for $B_1(\rho)$ and $B_2(\rho)$ 
\begin{eqnarray}\label{B1-B2}
&&B_1''(\rho)+{1 \over \rho}B_1'(\rho)-{3 \over \rho^2}B_1(\rho)=-{9 \over 32\rho}\\
&&B_2''(\rho)+{1 \over \rho}B_2'(\rho)-{7 \over \rho^2}B_2(\rho)=-{17 \over 32\rho}
\end{eqnarray}

which can be solved to give
\be\label{sol-d1}
B_1(\rho)=\frac{9}{64}\rho+c_1 \rho^{-\sqrt{3}}+c_2 \rho^{\sqrt{3}},\,\quad B_2(\rho)=\frac{17}{192}\rho+c'_1 \rho^{-\sqrt{7}}+c'_2 \rho^{\sqrt{7}}.
\ee
As above, we set the constants $c_1=c_0=0=c'_1=c'_2$ because the corresponding terms are either more singular than the leading term $\phi_0$ or produce no logarithms. As a result, one gets
\be\label{sol-d1-fin}
\varphi_2(\rho,\theta)=\frac{9}{64}\rho+\frac{17}{192} \rho \cos2\theta.
\ee

The careful reader may wonder why we stop the expansion of the solution at order $\frac{1}{a^2}$. This is so because for the calculation of the anomaly coefficient $d_1$ we need terms that scale as $\frac{1}{a^2}$ since for the cylinder 
\be\label{extrinsic}
Y^{\mu}_{ab}Y_{\mu}^{ab}=\frac{1}{2a^2}.
\ee
Moreover, we will argue in appendix \ref{Appendix:f2-f0^3} that the contribution involving \eqref{sol-d1-fin}  is zero. Consequently, the whole logarithmic contribution comes from $\varphi_0+\frac{1}{a}\varphi_1$ which we now calculate. 

As mentioned a few lines above, we ought to focus on the logarithmic term of the action which is proportional to $\frac{1}{a^2}$. This contribution will arise from the integrals very close to the defect, that is very close to $r=a$ and $x=0$.
By going back to the coordinates of \eqref{flat-metric-FT} which are $(x_0,r,\psi,x)$ we obtain
 \begin{equation} \label{ActionSYME-d1}
S_{\N = 4}^{(E)} = \frac{2}{g_{\text{\scalebox{.8}{YM}}}^2}\int [ dx_0 \, r d\psi ] \,\int dr  \,\int dx \ \text{tr}\bigg\{
\frac{1}{2} \left(D_{\mu}\varphi_{i+3}\right)^2 -\frac{1}{4}(\left[\varphi_{i+3},\varphi_{j+3}\right]^2)\bigg\} .
\end{equation}
The measure of \eqref{ActionSYME-d1} can be rewritten as  $r=a+r-a$ since the important part of the integration is around $r=a$. The reason for this will become apparent in a while. The integrand of the action functional can be evaluated to be
\begin{eqnarray}\label{Dsq}
&&\frac{1}{2} \text{tr}\left(D_{\mu}\varphi_{i+3}\right)^2=\frac{16 a^2+x^2}{32 a^2 \left((r-a)^2+x^2\right)^2}\frac{k(k^2-1)}{8} \approx \\
&&\frac{k(k^2-1)}{8\cdot 32 a^2}\frac{x}{(r-a)^2+x^2}\lim_{x \to 0} \frac{x}{(r-a)^2+x^2} +\cdots = \frac{k(k^2-1)}{8\cdot 32 a^2} \frac{1}{x} \pi \d(r-a)+\cdots\nonumber
\end{eqnarray}
To pass from the first to the second line of \eqref{Dsq} we have used the fact that we are very close to $x=0$ and approximate one of the fractions as a Dirac delta function. Then the integral with respect to $r$ becomes trivial while the integral with respect to $x$ will produce the logarithm we are after, i.e  $\int_\epsilon \frac{dx}{x}=-\ln \epsilon$.  The dots in \eqref{Dsq} indicate the term proportional to $16 a^2$ which does not give logarithms and does not have the correct ${1\over a^2}$ scaling.

The next logarithmic contribution originates form the $\varphi_{i+3}^4$ in \eqref{ActionSYME-d1}. 
\begin{eqnarray}\label{phi4}
-\frac{1}{4} \text{tr}\left[\varphi_{i+3},\varphi_{j+3}\right]^2=\frac{1}{4}\frac{k(k^2-1)}{8} \Big(\frac{6}{16 a^2}\frac{(r-a)^2}{((r-a)^2+x^2)^2}-\frac{4}{4 a}\frac{r-a}{((r-a)^2+x^2)^2}\Big)+\cdots\nonumber \\
%&&-\frac{k(k^2-1)}{32 a^2}\frac{10}{16}\frac{r-a}{(r-a)^2+x^2}\lim_{r \to a} \frac{r-a}{(r-a)^2+x^2} +\cdots =-\frac{k(k^2-1)}{32 a^2}\frac{10}{16} \pi \d(x) \frac{1}{r-a}+\cdots \nonumber
\end{eqnarray}
As above, the dots indicate terms that do not produce logarithmic terms. In \eqref{phi4} the term proportional to $\frac{6}{16 a^2}$ originates from keeping two factors of $\varphi_0$ and another two factors of $\varphi_1$ in the expansion of $(\varphi_0-\frac{1}{4a}\cos\theta)^4$ while the terms 
proportional to $\frac{4}{4 a}$ originates from keeping three factors of $\varphi_0$ and a single factor of $\varphi_1$ in the expansion of $(\varphi_0-\frac{1}{4a}\cos\theta)^4$. This second term will contribute because when inserted in the action it will be multiplied by $ \frac{r-a}{a}$ that comes from the second term in the measure $[dx_0 r d\psi]\approx a[dx_0 d\psi]+a [dx_0  \frac{r-a}{a} d\psi]$ along the defect. Taking the contribution of the measure into account the two terms in \eqref{phi4} become of the same order and result to 
\begin{eqnarray}\label{phi4-2}
-\frac{1}{4} \text{tr}\left[\varphi_{i+3},\varphi_{j+3}\right]^2\rightarrow %\frac{1}{4}\frac{k(k^2-1)}{8} \Big(\frac{6}{16 a^2}\frac{(r-a)^2}{((r-a)^2+x^2)^2}-\frac{4}{4 a}\frac{r-a}{((r-a)^2+x^2)^2}\Big)+\cdots\nonumber \\
&&-\frac{k(k^2-1)}{32 a^2}\frac{10}{16}\frac{r-a}{(r-a)^2+x^2}\lim_{r \to a} \frac{r-a}{(r-a)^2+x^2} +\cdots =\nonumber\\
&&=-\frac{k(k^2-1)}{32 a^2}\frac{10}{16} \pi \d(x) \frac{1}{r-a}+\cdots 
\end{eqnarray}
Now the $ \d(x)$ trivialises the $x$  integral while the logarithm will be produced by the integration with respect to $r$, i.e. $\int_{a+\epsilon} \frac{dr}{r-a}=-\ln \epsilon$.

Plugging both contributions \eqref{Dsq} and \eqref{phi4-2} in the action \eqref{ActionSYME-d1} we finally obtain
\begin{equation} \label{ActionSYME-d1-fin}
S_{\N = 4}^{(E)} = \frac{2}{g_{\text{\scalebox{.8}{YM}}}^2} \int [ dx_0 \, a d\psi ] \,\frac{k(k^2-1)}{8} \frac{\pi}{8 a^2}\, \ln\epsilon = \ln\epsilon \frac{2}{g_{\text{\scalebox{.8}{YM}}}^2}\int dx_0 d\psi \sqrt{\g}\,\frac{k(k^2-1)}{8}  \frac{2\pi}{8 }\,  Y^{\mu}_{ab}Y_{\mu}^{ab}.
\end{equation}
To derive the last expression we have used that for a cylinder of radius $a$ the induced metric is given by $\sqrt{\g}=a$ while 
$Y^{\mu}_{ab}Y_{\mu}^{ab}=\frac{1}{2 a^2}$.
By comparing the last equation to the generic expression \eqref{action-d1s} it is immediate to extract the anomaly coefficient $d_1$ at weak coupling to be 
\be\label{d1-weak-1}
d_1^{(\varphi_{4-6})}=24 \pi \frac{4 \pi}{8\,g_{\text{\scalebox{.8}{YM}}}^2}\,\frac{k(k^2-1)}{8} =\frac{3 \pi^2}{2\,g_{\text{\scalebox{.8}{YM}}}^2}\, k(k^2-1) >0\,.
\ee
 
 We now focus on the contribution of the three scalars $\varphi_i, \, i=1,2,3$ whose profiles is given in the second equation of \eqref{sol-2}. 
Since the three scalar commute with the remaining three, as well as among themselves, the field equation for the combination $\varPhi= \varphi_2+ i \varphi_3$ becomes
\be\label{radial-2}
\frac{\partial^2\varPhi}{\partial r^2}+\frac{1}{r}\frac{\partial \varPhi}{\partial r}+\frac{\partial^2\varPhi}{\partial x^2}= 0\, .
\ee
As above, we solve this  equation recursively in the limit where the defect is supported on a cylinder of very large radius $a$.
The solution can be written as 
\be\label{phi-exp-123}
\varPhi= \varPhi_0(\rho,\theta)+\frac{1}{a}\varPhi_1(\rho,\theta)+\frac{1}{a^2}\varPhi_2(\rho,\theta)\, .
\ee
$\bullet$ \underline{At order $a^0$}:\\
\be\label{a0123}
\frac{\partial^2\varPhi_0}{\partial r^2}+\frac{\partial^2\varPhi_0}{\partial x^2}= 0\Longrightarrow \varphi_0=\frac{C_1}{\rho e^{i \theta}}=C_1\frac{e^{-i \tan^{-1}\frac{x}{r-a}}}{\sqrt{(r-a)^2+x^2}}.
\ee
$\bullet$ \underline{At order $a^{-1}$}:\\
\be\label{a1123}
\frac{\partial^2\varPhi_1}{\partial r^2}+\frac{\partial^2\varPhi_0}{\partial x^2}+\frac{\partial \varPhi_0}{\partial r}= 0\Longrightarrow \frac{\partial^2\varPhi_1}{\partial \rho^2}+\frac{1}{\rho}\frac{\partial \varPhi_1}{\partial \rho}+\frac{1}{\rho^2}\frac{\partial^2\varPhi_1}{\partial \theta^2}=C_1\frac{ e^{-2 i \theta}}{\rho^2}\, ,
\ee
where $C_1=\sin\tilde\psi_0 \, C$ with $C$ being the diagonal matrix of \eqref{C-matr}.
Making now the ansatz $\varPhi_1=A(\rho) e^{-2 i \theta}$ one gets the ordinary differential equation
\be\label{ansatz-1}
A''(\rho )+\frac{A'(\rho )}{\rho }-\frac{4 A(\rho )}{\rho ^2}=\frac{C_1}{\rho ^2}\Longrightarrow A(\rho )=-{C_1 \over 4}+d_1 {1 \over \rho^2}+d_2 \rho^2\overset{d_1=d_2=0}{=}-{C_1 \over 4}
\ee
As a result we finally get 
\be\label{phi1-123}
\varPhi_1=-{C_1 \over 4} e^{-2 i  \theta}.
\ee\\
$\bullet$ \underline{At order $a^{-2}$}:\\
\be\label{a2123}
\frac{\partial^2\varPhi_2}{\partial r^2}+\frac{\partial^2\varPhi_2}{\partial x^2}-(r-a)\frac{\partial \varPhi_0}{\partial r}+\frac{\partial \varPhi_1}{\partial r}= 0\Longrightarrow \frac{\partial^2\varPhi_2}{\partial \rho^2}+\frac{1}{\rho}\frac{\partial \varPhi_2}{\partial \rho}+\frac{1}{\rho^2}\frac{\partial^2\varPhi_2}{\partial \theta^2}=-\frac{C_1}{4 \rho} (e^{- i \theta}+3 \,e^{- 3 i \theta})\, ,
\ee
To derive the last equality we have used the following expressions 
\begin{eqnarray}\label{fi0-fi1}
\frac{\partial \varPhi_0}{\partial r}=-C_1\frac{e^{-2 i \theta}}{\rho^2}, \qquad \frac{\partial \varPhi_1}{\partial r}=-C_1\frac{1}{4\, \rho}(e^{- i \theta}-e^{-3 i \theta}).
\end{eqnarray}
In order to solve \eqref{a2123} we make the ansatz 
\be\label{ans-1}
\varPhi_2=B_1(\rho) e^{-i \theta} +B_2(\rho) e^{-3 i \theta} 
\ee
which leads to the following ordinary differential equations for $B_1$ and $B_2$
\begin{eqnarray}\label{B1-B2}
&&B_1''(\rho )+\frac{B_1'(\rho )}{\rho }-\frac{ B_1}{\rho ^2}=-\frac{C_1}{4\rho }\Longrightarrow B_1(\rho )=\frac{C_1 \rho}{16}(1-2 \ln\rho)+\frac{d_1}{\rho}+d_2 \rho\\
&&B_2''(\rho )+\frac{B_2'(\rho )}{\rho }-\frac{ 9 B_2}{\rho ^2}=-\frac{3C_1}{4\rho }\Longrightarrow B_2(\rho )=\frac{3 C_1 \rho}{32}+\frac{d'_1}{\rho^3}+d'_2 \rho^3.
\end{eqnarray}

The relevant part of the action is written as \footnote{This is the case because $\varPhi$ commutes with itself, as well as with the other three scalars. }
 \begin{equation} \label{ActionSYME-d1-2}
S_{\N = 4}^{(E)} = \frac{2}{g_{\text{\scalebox{.8}{YM}}}^2}\int [ dx_0 \, r d\psi ] \,\int dr  \,\int dx  \frac{1}{2} \text{tr}
(D^{\mu}\bar \varPhi D_{\mu} \varPhi ), \quad{\rm where}\quad \varPhi= \varphi_2+ i \varphi_3.
\end{equation}
The action \eqref{ActionSYME-d1-2} may potentially receive  three contributions at the order $a^{-2}$ which is relevant for our calculation. The first originates when 
$ \varPhi_1$ is substituted in both  $ \varPhi$s of \eqref{ActionSYME-d1-2}. This contribution reads 
\begin{eqnarray}\label{S1}
S_1=\frac{2}{g_{\text{\scalebox{.8}{YM}}}^2}\int [ dx_0 \, a d\psi ] \,\int dr  \,\int dx \, {1 \over a^2} \frac{(\b^2+\g^2)\sin^2\tilde\psi_0 (N_1+k)}{4 \left((a-r)^2+x^2\right)}\, .
\end{eqnarray}
However, the integrals in the last equation does not produce any logarithms and as a result for our purposes one can safely set this contribution to zero, i.e. $S_1=0$.

The second potential contribution originates when 
$ \varPhi_0$ is substituted in one of the $\varPhi$s of \eqref{ActionSYME-d1-2} and $ \varPhi_2$ is substituted in the other. This contribution gives
\begin{eqnarray}\label{S2-Dmu}
(D^{\mu}\bar \varPhi_0 D_{\mu} \varPhi_2 +D^{\mu}\bar \varPhi_2 D_{\mu} \varPhi_0)=C_1^\dagger C_1\frac{5 (r-a)^2-x^2}{8 \left((a-r)^2+x^2\right)^2}\rightarrow C_1^\dagger C_1 {\pi \over 8} \Big(5 {\d(x)  \over r-a}-  {\d(r-a) \over x}\Big).\nonumber\\
\end{eqnarray}
By plugging the last equation in \eqref{ActionSYME-d1-2} and performing the integrals with respect to $x$ and $r$ we obtain
\begin{eqnarray}\label{S2}
S_2=-\frac{1}{g_{\text{\scalebox{.8}{YM}}}^2}\big((\b^2+\g^2)\sin^2\tilde\psi_0 (N_1+k)\big)\int [ dx_0 \, a d\psi ] \, \, {1 \over 2a^2} \pi \,\ln\epsilon .
\end{eqnarray}
To derive \eqref{S2} we have used that ${\rm tr} \,C_1^\dagger C_1= (\b^2+\g^2)\sin^2\tilde\psi_0 (N_1+k)$.
The last logarithmic contribution comes from the term
\begin{eqnarray}\label{S3-Dmu}
{r-a \over a}(D^{\mu}\bar \varPhi_0 D_{\mu} \varPhi_1 +D^{\mu}\bar \varPhi_1 D_{\mu} \varPhi_0)=C_1^\dagger C_1\frac{ (a-r) e^{-i \tan ^{-1}\left(\frac{x}{a-r}\right)}}{a \left((r-a)^2+x^2\right)^{3/2}}\rightarrow -C_1^\dagger C_1 {\pi \over a}   {\d(x)  \over r-a}.\nonumber\\
\end{eqnarray}
The term ${r-a \over a}$ in \eqref{S2-Dmu} comes from the measure $[ dx_0 \, r d\psi ]$ since $r$ can be written as $r=a(1+{r-a \over a})$. Overall the $a$ dependence is, as it should be ${1 \over a^2}$. One power of ${1\over a}$ comes fro the measure of the 2-dimensional defect and one power of ${1\over a}$ from $\varPhi_1$. 
Plugging \eqref{S3-Dmu} in the action we obtain
\begin{eqnarray}\label{S3}
S_3=\frac{1}{g_{\text{\scalebox{.8}{YM}}}^2}\big((\b^2+\g^2)\sin^2\tilde\psi_0 (N_1+k)\big)\int [ dx_0 \, a d\psi ] \,2{1 \over 2a^2} \pi \,\ln\epsilon .
\end{eqnarray}
Adding \eqref{S2} and \eqref{S3} we finally get
\begin{eqnarray}\label{S-final}
S_{\N = 4}^{(E)} =\frac{1}{g_{\text{\scalebox{.8}{YM}}}^2}\big((\b^2+\g^2)\sin^2\tilde\psi_0 (N_1+k)\big)\int [ dx_0 \, a d\psi ] \, {1 \over 2a^2} \pi \,\ln\epsilon =\nonumber\\ 
\frac{1}{g_{\text{\scalebox{.8}{YM}}}^2}\big((\b^2+\g^2)\sin^2\tilde\psi_0 (N_1+k)\big)\int [ dx_0 \, a d\psi ] \, Y^{\mu}_{ab}Y_{\mu}^{ab} \pi \,\ln\epsilon 
\end{eqnarray}
From the last equation it is straightforward to write down, by the use of \eqref{Wdiv} and \eqref{coeffs},  the contribution of the three scalars $\varphi_{1,2,3}$ to the anomaly coefficient $d_1$ at weak coupling. It reads
\be\label{d1-weak-2}
d_1^{(\varphi_{1-3})}=\frac{24 \pi^2}{g_{\text{\scalebox{.8}{YM}}}^2}\big((\b^2+\g^2)\sin^2\tilde\psi_0 (N_1+k)\big)>0\,.
\ee

We are now in position to write the final expression for the anomaly coefficient $d_1$ at weak coupling. This is obtained by adding \eqref{d1-weak-1} and \eqref{d1-weak-2}. 
\be\label{d1-weak-fin}
d_1^{(weak)}=\frac{3 \pi^2}{2\,g_{\text{\scalebox{.8}{YM}}}^2}\, k(k^2-1)+\frac{24 \pi^2}{g_{\text{\scalebox{.8}{YM}}}^2}\big((\b^2+\g^2)\sin^2\tilde\psi_0 (N_1+k)\big) >0\,.
\ee
Its value $d_1^{(weak)}$ is positive, as it should be for a unitary dCFT .

The behaviour of \eqref{d1-weak-fin} is in accordance with the interpolating structure of the corresponding dCFT of \cite{Georgiou:2025wbg}. At the end where $\cos\tilde\psi_0=0$, the first term in \eqref{d1-weak-fin} becomes zero because $k=0$ ((see discussion above equation (5.19)  of \cite{Georgiou:2025wbg})) and the weak coupling result for $d_1$ becomes in that limit 
\be\label{d1-weak-0}
d_1^{(weak)}(\tilde\psi_0=\pi/2)=\frac{24 \pi^2}{g_{\text{\scalebox{.8}{YM}}}^2}(\b^2+\g^2) N_1 \,.
\ee
%coefficient $d1$ vanishes, i.e. $b^{(weak)}(\tilde\psi_0=\pi/2)=0$, as it happens for the strong coupling result $b^{(strong)}(\tilde\psi_0=\pi/2)=0$. This vanishing is due to the corresponding vanishing of $k$ (see discussion above equation (5.19)  of \cite{Georgiou:2025wbg}). 
At the other end in which  $\cos\tilde\psi_0=1$,  it is the second term in \eqref{d1-weak-fin} which vanishes and one gets the weak coupling value of $d_1$ for the dCFT of \cite{Georgiou:2025mgg} which reads
\be\label{d1-weak-11}
d_1^{(weak)}(\tilde\psi_0=0)=\frac{3 \pi^2}{2\,g_{\text{\scalebox{.8}{YM}}}^2}\, k(k^2-1)\,.
\ee

\subsection{Comparison and agreement}\label{comp-3}

As for the case of the anomaly coefficient $b$ we compare the weak and strong coupling results for $d_1$ in the limit \eqref{dslimit}.
%where $\frac{k}{\sqrt{\l}}=\frac{\kappa}{\pi}\gg1\Leftrightarrow \s\rightarrow 0$ in which case $k$ can be approximated by $k=\frac{\sqrt{2 \l}}{\pi \s}$.

In this limit the weak coupling result \eqref{d1-weak-fin} becomes
\be\label{weak-d1-app}
d_1^{(weak)}=\frac{ \sqrt{\l} N}{\pi \s^3}\,\Big(3 \sqrt{2} -4 \sqrt{2}\,\bar\varepsilon-\frac{8 \sqrt{2}}{9 }\,\bar\varepsilon^2-\frac{160 \sqrt{2}}{243}\,\bar\varepsilon^3+{\mathcal O}(\bar\epsilon)^4)\Big)\, .
\ee
At this point, let us mention that the second term in the right hand side of \eqref{d1-weak-fin} should be completely ignored because it is suppressed in the limit $\s\rightarrow 0$ since its dependence on  $\s$ through the flux $k$ is of order $k \sin^2\tilde\psi_0\approx {\bar\epsilon\over \s}$ while the first term behaves as ${1\over \s^3}$.

In the same limit the strong coupling result for the anomaly coefficient $d_1$ \eqref{d1-strong-fin}  also becomes
\be\label{strong-d1-app}
d_1^{(strong)}=\frac{ \sqrt{\l} N}{\pi \s^3}\,\Big(3 \sqrt{2}+ \frac{4 \sqrt{2}}{3 }\,\bar\varepsilon-\frac{8 \sqrt{2}}{9 }\,\bar\varepsilon^2-\frac{32 \sqrt{2}}{27}\,\bar\varepsilon^3+{\mathcal O}(\bar\epsilon)^4)\Big)\, .
\ee
%In order to derive \eqref{strong-d1-app} we have used that in the aforementioned limit the condition $\s\rightarrow 0$ implies that $\rho\rightarrow 1$, $\tilde\psi_0\rightarrow 0$ and that $\kappa=\frac{\sqrt{2}}{\s}$, which is the same as $k=\frac{\sqrt{2 \l}}{\pi \s}$.
Comparing the last two equations we see that the agreement persists only for the leading order in the $\bar\varepsilon$ expansion. This is to be contrasted to the case of the anomaly coefficient $b$ for which the agreement holds to all orders in the $\bar\varepsilon$ expansion (see \eqref{agree}). The present author does not have a deeper understanding for this imbalance between the $b$ and $d_1$ anomaly coefficients.

In any case, the weak and strong coupling results agree to leading order and they read
\be\label{agree-d1}
d_1^{(strong)}=d_1^{(weak)}=3 \sqrt{2}\, \frac{ \sqrt{\l} N}{\pi \s^3}\, .
\ee
\section{Comparison with the literature}\label{Comp}
 
 The calculation of the defect Weyl anomaly coefficients $b$, $d_1$ and $d_2$  for the case of  interacting co-dimension 2 dCFTs is difficult and rare, especially if one is interested in {\it both} the weak and strong coupling regimes. One of these rare cases in which the values of the aforementioned central charges are known exactly is the case of the 1/2-BPS Gukov-Witten surface operators. For those operators the role of supersymmetry is instrumental since one can use localisation techniques in order to calculate the anomaly coefficients exactly \cite{Chalabi:2020iie}.
 
 Our case is completely different since our dualities do not preserve any amount of the supersymmetry which the theory  without the defect has. However, in a certain limit, namely when $\tilde\psi_0=0$, our D5 probe brane becomes  an 1/2-BPS object preserving, thus, half of the Poincare supersymmetries, as well as half of the conformal supersymmetries (see \cite{Georgiou:2025wbg}).  Thus, it would be interesting to check whether our results comply with the exact results for the Gukov-Witten surface operators. This would be a strong consistency check  of our results especially since our calculation is unique in the sense that no similar calculation has been ever performed in the context of non-supersymmetric defect CFTs.
 
 Let us start with the simplest case of the anomaly coefficient multiplying the extrinsic curvature term, i.e. $d_1$. For the general D5 brane of \cite{Georgiou:2025wbg} its value has been calculated in section \ref{weak-3} and is given by equation \eqref{d1-weak-fin}. In the limit $\tilde\psi_0=0$, one obtains \eqref{d1-weak-0} which we rewrite here for the convenience of the reader
 \be\label{d1-weak-01}
d_1^{(weak)}(\tilde\psi_0=\pi/2)=\frac{24 \pi^2}{g_{\text{\scalebox{.8}{YM}}}^2}(\b^2+\g^2) N_1 =\frac{24 \pi^2}{\lambda}(\b^2+\g^2) N N_1\,.
\ee
The exact result for the Gukov-Witten surface operators can be found, for example, in equation (2.5b) of \cite{Chalabi:2020iie} and reads
  \be\label{d2-exact}
d_2=d_1=3(N^2-(N-N_1)^2-N_1^2)+\frac{24 \pi^2}{g_{\text{\scalebox{.8}{YM}}}^2}(\b^2+\g^2) N_1 =6 N N_1+\frac{24 \pi^2}{\lambda}(\b^2+\g^2) N N_1\,.
\ee
In \eqref{d2-exact} we have taken into account that for a supersymmetric defect $d_2=d_1$ and that our configuration comprises of two sets of branes. $N_1$ coincident ones with $\beta_i=\beta$ and   $\gamma_i=\gamma$ where $i=1,\ldots, N_1$ and $N-N_1$ ones with $\beta_i=0=\gamma_i,\,  i=1,\ldots N-N_1$. Thus we see that our result is in complete agreement with the exact result \eqref{d2-exact}. Needless to say, the first term, $6 N N_1$, in \eqref{d2-exact} is subleading in the small $\lambda$ expansion relevant to our result. In order to reproduce this subleading term, one should include the one-loop correction to \eqref{d1-weak-fin}. It would be interesting to perform this calculation. Furthermore, we see that our result is consistent with the fact that for a supersymmeteic defect $d_2=d_1$ \cite{Bianchi:2019sxz}.

We now turn to the anomaly coefficient $b$. Here the situation is more intricate due to the fact that, in the corresponding limit $\tilde\psi_0=\pi/2$, our D5 brane configuration becomes singular since the $S^2\subset S^5$ sphere, which the D5 brane wraps, shrinks to zero size. This can be seen from the induced on the brane metric given by equation (2.13) of \cite{Georgiou:2025wbg}. From this equation it can easily seen that the radius of $S^2$, which is 
$\cos\tilde\psi_0$, becomes zero at $\tilde\psi_0=\pi/2$. As a consequence, in this limit the action \eqref{action-on-shell} of our D5 brane goes to zero because the action is proportional to the volume of $S^2$. This vanishing of the action results to the vanishing of the coefficient $b$ in the limit $\tilde\psi_0=\pi/2$, see \eqref{b-2} and the discussion below it.\footnote{The same reasoning holds for the strong coupling result for $d_1$ presented in section \ref{strong-3}.}

In order to make contact to the exact value for the $b$ coefficient for the Gukov-Witten operators 
\be\label{b-exact}
b=3(N^2-(N-N_1)^2-N_1^2)=6 N N_1\,.
\ee
presented in equation (2.5a) of \cite{Chalabi:2020iie},
one should somehow factor out the zero coming for the shrinking of the $S^2$. This can be done by dividing the action, and as a consequence the result for the $b$ coefficient, by the vanishing flux through the two-sphere $S^2$, namely by 
\begin{equation}\label{kappa-1-1}
k=\int_{S^2} \frac{F}{2 \, \pi} \quad  \Rightarrow \quad k=\frac{\kappa}{\pi \a'}=\frac{\kappa \sqrt{\lambda}}{\pi}\, \overset{\tilde\psi_0\rightarrow \pi/2}{=}\frac{ \sqrt{\lambda}}{\pi}\, \cos\tilde\psi_0 \frac{\sqrt{2(1+\s^2)}}{\s}.
\end{equation}
Here we have used \eqref{kappa-2} for $\kappa$. 
Now by dividing \eqref{b-2} by $k$ and by taking into account that in the limit $\tilde\psi_0\rightarrow \pi/2$, $\rho\rightarrow 1$ (see discussion around \eqref{boundss}), one obtains the 
correct $\lambda$-independent result 
\be\label{b-ours}
b=6 N\,.
\ee
This is precisely the exact result for $b$ of equation \eqref{b-exact} when multiplied by $N_1$  which is the number of the D5 probe branes.\footnote{The calculation of $b$ in  \eqref{b-2} has been done for a single probe brane.}

We conclude that our results for the Weyl anomaly coefficients are consistent with the corresponding values of the anomaly coefficients for the special case of the Gukov-Witten surface operators.

\section{Conclusions}\label{concl}
In this work, we have computed the Weyl anomaly coefficients associated with the co-dimension two non-supersymmetric defect CFTs that were introduced in the works \cite{Georgiou:2025mgg}  and  \cite{Georgiou:2025wbg}, at both the weak and strong coupling regimes. In particular, we firstly determined the type-A Weyl anomaly coefficient $b$ associated with the intrinsic scalar curvature of the defect. At strong coupling, we employed the dual D5-brane solutions in Euclidean signature, where the defect is supported on an $S^2$ submanifold of the Euclidean $AdS_3\times S^1$ boundary. At weak coupling, we used the classical solutions of the ${\cal N}=4$ SYM equations of motion, previously conjectured to describe the defects dual to the D5-brane configurations. 

Subsequently, we evaluated the type-B Weyl anomaly coefficient $d_1$ associated with the defect's  extrinsic curvature, first at strong coupling and subsequently at weak coupling. For the holographic calculation one needs a D5 brane configuration with non-zero extrinsic curvature. We found the behaviour of such a configuration near the boundary by considering small perturbations of the flat co-dimension 2 defect and determining how these propagate into the bulk.  At weak coupling, we considered a cylindrical defect of large radius $a$ in flat Euclidean space and solved the ${\cal N}=4$ SYM equations perturbatively in $1/a$, allowing us to extract the logarithmic contribution to the action controlled by the extrinsic curvature.

In a well-defined limit, the weak- and strong-coupling results for both $b$ and $d_1$ exhibit an impressive agreement. This provides non-trivial evidence for the proposed holographic dualities of \cite{Georgiou:2025mgg,Georgiou:2025wbg} and supports the interpolating structure identified in \cite{Georgiou:2025wbg}. Notably, the coefficient $b$ is found to become negative in a finite region of parameter space, passing through zero and then becoming positive as one of the parameters vary. To our knowledge, this constitutes the first explicit example of an interacting dCFT with negative $b$. By contrast, $d_1$ remains positive throughout, in agreement with unitarity. Let us mention that, although there is a monotonicity $b$-theorem \cite{Jensen:2015swa}, this constrains differences along RG flows, namely that $b_{UV}\ge b_{IR}$, and not the absolute value of $b$.

Several directions merit further investigation. In particular, it would be important to compute one-loop corrections to $b$ at both strong and weak coupling, extending the analysis of \cite{Jiang:2024wzs}. It would also be natural to study higher-point correlation functions in this framework, along the lines of \cite{Georgiou:2023yak,Linardopoulos:2026mut}.

\subsection*{Acknowledgements}
I wish to thank Georgios Linardopoulos  and Dimitrios Zoakos  for  discussions on various aspects of defect CFTs.

%\newpage

\appendix

\section{Matrix relations}
\label{Appendix:details_matrices}
In this appendix we gather some useful relations regarding the matrices defining the vevs of the scalar fields.
In section \ref{weak-1} we follow the conventions of \cite{Georgiou:2025mgg}. In particular, the matrices $t_i$ appearing in \eqref{vevs-1} 
 realise a $k$-dimensional irreducible representation of $\mathfrak{su}\left(2\right)$ and satisfy
\begin{IEEEeqnarray}{c}\label{tmatr}
\left[t_i, t_j\right] = \frac{i}{ \sqrt{2}}\epsilon_{ijl}t_l, \qquad i,j,l = 1,2,3.
\end{IEEEeqnarray}
Furthermore, the matrices $t_i$ also satisfy 
\be\label{comm}
t_i=\sum_{j=1}^3\left[t_j,\left[t_j,t_i\right]\right]
\ee
which is fully consistent with \eqref{tmatr}.
Let us mention that in the dual gravity description, $k$ corresponds to the integer flux through the $S^2\subset S^5$ (see \eqref{kappa}).
We will also need the following useful relation of the matrices $t_i$ 
\be\label{normm}
\sum_{i=1}^3 (\varphi_{i+3}^{\text{cl}})^2=\frac{1}{8 r'^2}(k^2-1) {\mathbf 1}_{k\times k} \oplus {\mathbf 0}_{(N-k)\times (N-k) }.
\ee
By using \eqref{tmatr} and \eqref{normm}  it is straightforward to show that 
\be\label{comm}
\text{tr}[\varphi_i, \varphi_j]^2=-\sum_{i=1}^3\text{tr}\, \varphi_i^2=-\frac{k(k^2-1)}{r'^2\, 8}
\ee 
\be\label{comm-1}
\text{tr}[t_i, t_j]^2=-\sum_{i=1}^3\text{tr}\, t_i^2=-\frac{k(k^2-1)}{8}
\ee 
All the equations for the scalar fields hold when the defect is planar embedded in flat spacetime. When the field theory lives in $AdS_3\times S^1$ the $r'^2$ in the denominators is absent.
\section{ Vanishing $\varphi_2$ contribution}
\label{Appendix:f2-f0^3}
In this appendix we show that the ${1\over a^2}$ contribution to \eqref{ActionSYME-d1} coming from the correction $\varphi_2$  vanishes.
Indeed,  one has 
$\frac{1}{2} \left(D_{\mu}\varphi_{i+3}\right)^2 -\frac{1}{4}(\left[\varphi_{i+3},\varphi_{j+3}\right]^2)$
\begin{eqnarray}\label{DD}
\frac{1}{2} \text{tr}\,\left(D_{\mu}\varphi_{i+3}\right)^2= D_{\mu}\varphi_0 D^{\mu}\varphi_2\sum_{i=1}^3\text{tr}\, t_i^2=-\frac{17 \cos \left(2 \tan ^{-1}\left(\frac{x}{a-r}\right)\right)+27}{192 \left((a-r)^2+x^2\right)}\sum_{i=1}^3\text{tr}\, t_i^2
\end{eqnarray}

\begin{eqnarray}\label{f^4}
-\frac{1}{4}\text{tr}\,(\left[\varphi_{i+3},\varphi_{j+3}\right]^2)=-\frac{1}{4} 4 \,\varphi_0^3\, \varphi_2 \,\text{tr}[t_i, t_j]^2=-\frac{17 \cos \left(2 \tan ^{-1}\left(\frac{x}{a-r}\right)\right)+27}{192 \left((a-r)^2+x^2\right)}\,\text{tr}[t_i, t_j]^2\nonumber\\
\end{eqnarray}
By the use of \eqref{comm-1} it is obvious that the sum of \eqref{DD} and \eqref{f^4} is zero proving the claim the correction involving $\varphi_2$ gives zero contribution at the relevant order ${1 \over a^2}$.

\bibliographystyle{utphys}

\bibliography{refs}

\end{document}